# Theory of plasmon-enhanced Förster energy transfer in optically-excited semiconductor and metal nanoparticles

Alexander O. Govorov[1]*, Jaebeom Lee[2,3], and Nicholas A. Kotov[2]

[1]*Department of Physics and Astronomy, Ohio University, Athens, OH, 45701*

[2]*Department of Chemical Engineering, Biomedical Engineering, and Material Sciences and Engineering University of Michigan, Ann Arbor, MI, 4810*

[3]*Pusan National University, Busan 609-735, Korea*

## ABSTRACT

We describe the process of Förster transfer between semiconductor nanoparticles in the presence of a metal subsystem (metal nanocrystals).   In the presence of metal nanocrystals, the Förster process can become faster and more long-range. The enhancement of Förster transfer occurs due to the effect of plasmon-assisted amplification of electric fields inside the nanoscale assembly. Simultaneously, metal nanocrystals lead to an increase of energy losses during the Förster transfer process. We derive convenient equations for the energy transfer rates, photoluminescence intensities, and energy dissipation rates in the please of plasmon resonances. Because of strong dissipation due to the metal, an experimental observation of plasmon-enhanced Förster transfer requires special conditions. As possible experimental methods, we consider cw- and time-resolved photoluminescence studies and describe the conditions to observe plasmon-enhanced transfer.  In particular, we show that the photoluminescence spectra should be carefully analyzed since the plasmon-enhanced Förster effect can appear together with strong exciton energy dissipation. Our results can be applied to a variety of experimental nanoscale systems.

*Corresponding Author: Govorov@ohiou.edu

**Introduction**

In many experiments, colloidal nanoparticles (NPs) strongly confine carriers and do not permit efficient tunnel coupling. However, instead of direct tunnel coupling, the colloidal NPs permit **Förster transfer (FT)** of optically-generated excitons.[1] The FT mechanism comes from the inter-NP Coulomb interaction and does not require the tunnel coupling between semiconductor nanoparticles (SNPs). The FT process can also be viewed as exciton energy flow a donor to accepter nanocrystal/molecule. Fig. 1a illustrates a process of exciton transfer from a donor SNP1 to acceptor SNP2. Both SNPs can emit photons and FT can be observed experimentally as enhanced emission from SNP2.[2,3,4] Another type of structures studied in recent experiments consists of SNPs and metal nanocrystals. In these structures, individual NPs can also interact via the Coulomb forces. However, the character of interaction can be different due to large electric dipole moments and plasmon resonances in metal nanoparticles (MNPs). The interaction between excitons in SNPs and surface plasmons in MNPs can result in enhanced exciton emission due to the so-called plasmon-induced field enhancement effect.[5,6,7,8] Such plasmon-enhanced exciton emission can appear due to amplification of both absorption and emission processes.[8] Simultaneously, in the metal-semiconductor structures, the exciton energy can be transferred from SNPs to MNPs and then converted into heat. SNP-MNP energy transfer becomes especially strong in the exciton-plasmon resonance and can be observed as shortening of the exciton lifetime.[5,8,9] The FT process can also be combined with the plasmon resonances in the structures incorporating a emitter (dye or SNP) and MNPs. For the case of dye molecules and silver MNPs, such plasmon-assisted FT process was studied experimentally in ref.[10] and theoretically in ref.[11].

Here we study theoretically the FT process between SNPs in the presence of metal nanocrystals. Our calculations show an effect of long-range FT assisted by plasmons. Simultaneously, the exciton lifetime can become shorter due to energy transfer to the metal component. This leads to energy losses and can strongly reduce the efficiency of FT process. Our formalism is based on the rate equations and fluctuation-dissipation theorem. Previously, plasmon-assisted FT was studies for dye molecules and silver NPs.[10,11] Here we describe the plasmon-assisted FT process for the case of SNPs. From the point of view of physics, SNPs and dye molecules have important differences: 1) A SNP has three optically active excitons which take part in the FT process; it also has several dark excitons. At the same time, a dye molecule can be well approximated as a single optically-active dipole. 2) Because of fast spin-flips at room temperature, dark and bright excitons in a SNP become almost equally populated. The resultant FTR rate should incorporate the exciton populations in a SNP. For the case of a dye molecule, such exciton dynamics is not involved. 3) SNPs have a large background dielectric constant (about 5-10). This background dielectric constant strongly affects the process. From the mathematical side, dye molecules and SNPs should also be treated differently. In ref.[11], dye molecule was treated as a point-like dipole in the vicinity of metal particle. For this case, the FT rate can be written analytically as an infinite sum. A SNP is not a point-like dipole since it size can be comparable with the inter-nanocrystal distances in the complex. In addition, SNP has a large background dielectric constant and, therefore, the effect of surface charges on the electrostatics and FT process is strong. Therefore, the FT process rate should be modeled numerically. Here we suggest a convenient numerical formalism to compute the FT rates for complexes with arbitrary architecture. Our formalism is based on the multi-pole expansion and fluctuation-dissipation theorem. In addition, we derive convenient analytical equations for the dipole limit; these analytical expressions can be used to obtain reliable estimates. For the general case, the FT and energy dissipation rates should be calculated numerically. This paper considers several material systems where plasmon-enhanced FT can be observed. Our results suggest that the conditions to observe long-range FT are quite peculiar due to strong energy dissipation in MNPs. In the paper, we mostly focus on the optical effects at room temperature.

## 1. Rate equations

Here we describe energy transfer between two SNPs in the presence of metal component (Fig. 1a). In our system, SNPs have different optical band gaps due to the size-quantization. A small nanoparticle (SNP1) absorbs an incident photon and then acts as a donor supplying an exciton to a larger nanoparticle (SNP2) which represents an acceptor. Exciton transfer between the donor and acceptor is induced by the Coulomb interaction. This transfer process has three steps: fast energy relaxation in SNP1, FT process, and fast energy relaxation in SNP2 (Fig. 1b). Finally, SNP2 emits a secondary photon at a lower energy. The FT scheme shown in Fig. 1b is rather conventional[1,2]; this scheme also includes direct absorption of incident photons by the SNP2. In the presence of efficient

FT, optical emission of SNP2 should be much stronger than that of SNP1 because of directional exciton flow from SNP1 to SNP2. The MNPs in our scheme can strongly change the FT probability and energy dissipation rates. Depending on the parameters of the system, the FT rate can be enhanced or suppressed in the presence of MNPs.

Below, we consider the following conditions: a NP complex size ($L_{complex}$) is much smaller than the wavelength of incident light ($\lambda_{laser}$), i.e. $\lambda_{laser} \gg L_{complex}$. Under the typical experimental conditions, we also have $\lambda_{SNP1} \sim \lambda_{SNP2} \sim \lambda_{laser}$ and $\lambda_{SNP1}, \lambda_{SNP2} \gg L_{complex}$, where $\lambda_{SNP1}$ and $\lambda_{SNP2}$ are the exciton emission wavelengths of SNP1 and SNP2, respectively. In addition, we assume that intra-band relaxation processes within SNPs are fast (continuous blue arrows in Fig.1b) whereas the exciton recombination and FT processes are slower. Typically, the intra-band energy relaxation time $\tau_{ener}$ is in the range of $10\,ps$ while the recombination and FT times ($\tau_{rec}$ and $\tau_{transfer}$) are $10-20\,ns$[2,4] at room temperature. Another fast process involved in our scheme is plasmon relaxation in MNP; the lifetime of plasmon ($\tau_{plas}$) is very short, in the range of $fs$.

Excitons in their excited states ($|exc_1\rangle$ and $|exc_2\rangle$) can be created optically or due to FT (Fig. 1b). After fast intra-band relaxation, optically-generated excitons in SNP1 and SNP2 reside in the low-energy exciton states, $|\psi_1\rangle$ and $|\psi_2\rangle$ (see Fig. 1b). These exciton states with low energies ($|\psi_1\rangle$ and $|\psi_2\rangle$) will be denoted as $|i\alpha\rangle$, where $\alpha = 1,2,...8$ and $i = 1,2$ is the SNP number. The energies of the states $|\psi_i\rangle$ with $i = 1,2$ will be denoted as $E_1 = \hbar\omega_1$ and $E_2 = \hbar\omega_2$, respectively; $\omega_i = 2\pi c/\lambda_{SNPi}$. The number of excitons with low energies in SNP is eight since there are four single-electron states in the valence band (heavy and light holes with $j_z = \pm 1/2, \pm 3/2$) and two electrons states in the conduction band (with spins $s_z = \pm 1/2$).[12,13] The total recombination rate of exciton $|i\alpha\rangle$ is given by

$$\gamma_{exc,i\alpha} = \gamma_{rad,i\alpha} + \gamma^0_{non-rad,i\alpha} + \gamma_{metal,i\alpha},$$

where $\gamma_{rad,i\alpha}$ and $\gamma^0_{non-rad,i\alpha}$ are radiative and non-radiative recombination rates, respectively. Here we assume, for simplicity, that all "intrinsic" non-radiative rates are the same: $\gamma^0_{non-rad,i\alpha} = \gamma^0_{non-rad}$; $\gamma_{metal,i\alpha}$ are the rates of exciton-energy transfer from SNPs to MNPs. For fast intra-band energy relaxation of excitons, the rate equations at room temperature take the form:

$$\frac{dn_{1\alpha}}{dt} = I_{1\alpha} - \left(\gamma^0_{non-rad} + \gamma_{rad,1\alpha} + \gamma_{metal,1\alpha} + \sum_{\alpha''=x,y,z}\gamma_{Forster,1\alpha\to 2\alpha''}\right)\cdot n_{1\alpha} - \gamma_{spin}\sum_{\alpha''\neq\alpha}(n_{2\alpha} - n_{2\alpha''})$$

$$\frac{dn_{2\alpha'}}{dt} = I_{2\alpha'} - (\gamma^0_{non-rad} + \gamma_{rad,2\alpha} + \gamma_{metal,2\alpha'})\cdot n_{2\alpha'} + \sum_{\alpha''=x,y,z}\gamma_{Forster,1\alpha''\to 2\alpha'}\cdot n_{1\alpha''} - \gamma_{spin}\sum_{\alpha'\neq\alpha''}(n_{2\alpha'} - n_{2\alpha''}),$$

(1)

where $n_{1\alpha}$ and $n_{2\alpha'}$ are averaged numbers of excitons in the low-energy excitons states $|1\alpha\rangle$ and $|2\alpha'\rangle$, respectively. The exciton-state index $\alpha$ and $\alpha'$ vary independently ($\alpha = 1,2,...8$ and $\alpha' = 1,2,...8$). The rates $I_{i\alpha}$ describe optical generation of excitons in SNPs in the presence of laser light. The rate $\gamma_{Forster,1\alpha\to 2\alpha'}$ represents the unidirectional FT process $1\alpha \to 2\alpha'$. The FT process is unidirectional because of fast energy relaxation and trapping of excitons in SNP2. The rate $\gamma_{spin}$ is responsible for spin relaxation between different exciton ground states. We should stress that the equations (1) are given

for the room temperature regime. At room temperature, spin relaxation is typically fast (in the *ps* range) and we obtain $n_{i\alpha} \approx n_{i\alpha'}$. At low temperatures, the spin relaxation can be slow and therefore $n_{i\alpha} \neq n_{i\alpha'}$ for $\alpha \neq \alpha'$.[14] Below we will comment more on the low-temperature regime.

Using the approximation of fast spin relaxation ($n_{i\alpha} \approx n_{i\alpha'}$, $\gamma_{spin} \to \infty$) and the steady-state condition, we can easily solve eqs. 1 for the total exciton populations in SNPs:

$$n_1 = \sum_\alpha n_{1\alpha} = \frac{8\sum_\alpha I_{1\alpha}}{\sum_\alpha \gamma^0_{non-rad} + \gamma_{rad,1\alpha} + \gamma_{metal,1\alpha} + \sum_{\alpha'\alpha''} \gamma_{Forster,1\alpha \to 2\alpha''}} = \frac{I_{1,tot}}{\gamma_1},$$

$$n_2 = \sum_\alpha n_{2\alpha} = 8\frac{\sum_\beta I_{2\alpha} + \frac{n_1}{8}\sum_{\alpha',\alpha''}\gamma_{Forster,1\alpha' \to 2\alpha''}}{\sum_\alpha \gamma^0_{non-rad} + \gamma_{rad,2\alpha} + \gamma_{metal,1\alpha}} = \frac{I_{2,tot} + n_1 \cdot \gamma_{Forster}}{\gamma_2}, \quad (2)$$

where $I_{i,tot} = \sum_\alpha I_{i\alpha}$ are the total absorption rates in SNPs. The rates $\gamma_i$ and $\gamma_{Forster}$ are the averaged recombination rates of excitons and the averaged FT rate, respectively. These rates are given by:

$$\gamma_1 = \sum_\alpha \left( \gamma^0_{non-rad} + \gamma_{rad,1\alpha} + \gamma_{metal,1\alpha} + \sum_{\alpha''} \gamma_{Forster,1\alpha \to 2\alpha''} \right)/8,$$

$$\gamma_2 = \left( \sum_\alpha \gamma^0_{non-rad} + \gamma_{rad,2\alpha} + \gamma_{metal,1\alpha} \right)/8,$$

$$\gamma_{Forster} = \sum_{\alpha'\alpha''} \gamma_{Forster,1\alpha \to 2\alpha''}/8.$$

The exciton rates depend on the exciton wave functions. For the sake of simplicity, we employ here simplified wave functions without the mixing between heavy- and light-hole states. In other words, we use the approximation: $\Psi_{exc} = \Psi_e(\mathbf{r}_e) u_{s_z}(\mathbf{r}_e) \Psi_h(\mathbf{r}_h) u_{j_z}(\mathbf{r}_h)$, where $u_{s_z}(\mathbf{r}_e)$ and $u_{j_z}(\mathbf{r}_h)$ are the Bloch wave functions in the conduction and valence bands, respectively. In the absence of valence-band mixing, the corresponding envelope functions for a spherical SNP are $\Psi_e = \Psi_h = \Psi = \sin(\pi r / R_{SNP})/(\sqrt{2 R_{SNP} \pi} \cdot r)$, where $R_{SNP}$ is a radius of SNP. Among the functions $|\alpha\rangle = \Psi_e(\mathbf{r}_e) u_{S_z}(\mathbf{r}_e) \Psi_h(\mathbf{r}_h) u_{j_z}(\mathbf{r}_h)$, there are several optically-dark states. We can make linear combinations of these functions and obtain a more convenient set of states $|new\rangle = \Psi_e(\mathbf{r}_e) \Psi_h(\mathbf{r}_h) u_{new}(\mathbf{r}_h, \mathbf{r}_e)$. Among the new functions, there are three optically-active states ($\beta = x, y, z$) and five dark states ($\nu = 1,2,...5$). The $\beta$-exciton has optical dipole moment in the $\beta$-direction. We can make this convenient choice of wavefunctions due to the spherical symmetry of SNPs.[15] Therefore, we can write the transfer rates in terms of the bright states:

$$\gamma_1 = \gamma_{non-rad}^0 + \sum_{\beta = x,y,x} \left( \gamma_{rad,1\beta} + \gamma_{metal,1\beta} + \sum_{\beta''} \gamma_{Forster,1\beta \to 2\beta''} \right)/8,$$

$$\gamma_2 = \gamma_{non-rad}^0 + \left( \sum_{\beta} \gamma_{rad,2\beta} + \gamma_{metal,1\beta} \right)/8, \quad \gamma_{Forster} = \sum_{\beta'\beta''} \gamma_{Forster,1\beta \to 2\beta''}/8. \quad (3)$$

In our model, the dark states do not take part in optical and energy-transfer processes. However, they take part in exciton dynamics due to spin flips and this brings the factor 1/8 in eq. 3. We should note that the above approximation makes our description much more convenient and transparent, but it ignores the valence-band mixing. The valence-band mixing effect can lead to non-zero optical and FT matrix elements for dark excitons if the size of SNPs is comparable with inter-NP distance. This is due to inhomogeneous fields inside a nano-complex induced by the long-wavelength photonic fields ($\lambda \gg L_{complex}$). Numerically, the valence-band mixing effects in the optical and FT matrix elements are relatively small.[16] We also should mention that the above simplification holds for small SNPs ($R_{SNP} \ll L_{complex}$) even in the presence of the valance-band mixing effect.

We also should mention the effect of energy splitting between bright and dark excitons and the influence of temperature (*T*). The darks states in nanocrystals have typically a lower energy. The dark-bright exciton splitting $\Delta_{exc}$ is ~ $5 meV$ for both CdTe and CdSe SNPs with $R_{SNP} \sim 2 mn$.[3,13] This is essentially smaller than the thermal energy at room temperature: $k_B T \approx 26 meV$. Therefore, our assumption $n_{i\alpha} \approx n_{i\alpha'}$ is a good approximation at room temperature. At low temperatures, the spin-flip rates for the processes *dark→bright* should have an exponential factor $e^{-\Delta_{exc}/k_B T}$. Therefore, at low temperatures, the dark-exciton population should be larger than the bright-exciton one. Simultaneously, the averaged energy-transfer rates will also acquire the same Boltzmann factor $e^{-\Delta_{exc}/k_B T}$ and all inter-NP transfer processes should slow down as temperature decreases. Overall, the exciton dynamics at low *T* can become more complicated; this comes also from the fact that the processes *dark→bright* become slow and bring additional exponential functions to photoluminescence kinetics.[17] Some of these issues were addressed in the recent experimental papers.[3,17]

The parameters $I_{i\alpha}$ and $\gamma_{rad,i\beta}$ can be strongly modified inside a NP complex due to the plasmon resonances in MNPs. To account for this effect, we will calculate these parameters in the following way[18]:

$$\gamma_{rad,i\beta} = P^a_{i\beta}(\omega_i) \cdot \gamma^0_{rad,exc}, \quad I_{i,tot} = P^b_i(\omega_{laser}) \cdot I^0_i, \quad (4)$$

where $\omega_i = E_i/\hbar$ is the optical emission frequency of $i$-SNP, $\omega_{laser}$ is the exciting laser frequency, $\gamma^0_{rad,exc}$ is the radiative rate of SNP in the absence of the metal; for simplicity, we assume $\gamma^0_{rad,exc}$ is the same for both SNPs. The enhancement factor $P^a_{i\beta}(\omega_i)$ is introduced for the process of emission from the lowest exciton states:

$$P^a_{i\beta}(\omega) = \frac{\left|\int_{V_i} \vec{E}_{actual} \vec{D}^a_{i\beta}(\vec{r}) dV\right|^2}{\left|\int_{V_i} \vec{E}_{no\ metal} \vec{D}^a_{i\beta}(\vec{r}) dV\right|^2}, \quad (5)$$

where $\vec{E}_{actual}$ is the resultant electric field inside the SNP induced by the external field $\vec{E}_0(t) = \vec{E}_0 \cdot e^{-i\omega t}$ ($\lambda_{SNPi} \gg R_{SNPi}, L_{complex}$); the function $\vec{D}^a_{i\beta}(\vec{r})$ describes the spatial distribution of inter-band dipole moment in the ground-state exciton $|i\beta\rangle$. For our choice of wavefunctions, we obtain $\vec{D}^a_{i\beta}(\vec{r}) = \vec{e}_\beta \cdot d_{exc} \Psi^2(\vec{r} - \vec{r}_{SNPi})$, where $\vec{r}_{SNPi}$ is the position of $i$-SNP, $d_{exc} = \langle 0|x|ix\rangle$ is the inter-band dipole moment of SNP ($d_{exc}$ is few Å typically), $\vec{e}_\beta$ is the unit vector ($\vec{x}$, $\vec{y}$, or $\vec{z}$) parallel to the $\beta$-direction, and $\beta = x, y, z$. The integral $\int_{V_i} \vec{E}_{no\ metal} \vec{D}^a_{i\beta}(\vec{r}) dV = 3 d_{exc} E_{0,\beta} \varepsilon_0/(2\varepsilon_0 + \varepsilon_s)$, where $\varepsilon_0$ and $\varepsilon_s$ are the dielectric constants of the surrounding media and semiconductor, respectively. The factor $R = 3\varepsilon_0 / (2\varepsilon_0 + \varepsilon_s)$ appears due to screening of external field inside a dielectric sphere.[19]

The absorption process is described with a similar factor:

$$P_i^b(\omega) = \frac{\sum_{exc_i} \left| \int_{Vi} \vec{E}_{actual,laser} \vec{D}_{exc_i}^b(\vec{r}) dV \right|^2}{\left| \sum_{exc_i} \int_{Vi} \vec{E}_{no\,metal,laser} \vec{D}_{exc_i}^b(\vec{r}) dV \right|^2}, \qquad (6)$$

where $\vec{E}_{no\,metal,laser}(t) = \vec{E}_0 \cdot e^{-i\omega t}$ is the external laser field, $|exc_i\rangle$ are the excited states of exciton in $i$-SNP. In this case, the function $\vec{D}_{exc_i}^b(\vec{r})$ describes the local dipole moment of an excited state of exciton. Since excited states have a more uniform spatial probability distribution, we will assume that the local dipole moment inside a SNP is a constant: $\vec{D}_{exc_i}^b(\vec{r}) = \vec{e}_{exc_i} \cdot d_{exc} / V_{SNPi}$ for $|\vec{r} - \vec{r}_{SNP2}| < R_{SNP2}$, where $V_{SNPi}$ is the volume of $i$-SNP. For bright excited states, $|exc_i\rangle = |\gamma i\rangle$, where $\gamma = x, y, z$; $\vec{e}_{exc_i} = \vec{e}_\gamma$ are the unit vectors. Then, the denominator of eq. 6 can be written as $\left| \sum_\gamma \int_{Vi} \vec{E}_{no\,metal} \vec{D}_{exc_i}^b(\vec{r}) dV \right|^2 = d_{exc}^2 R^2 E_0^2$. Note that, in a NP complex, the factor (6) depends on the direction of the laser field $\vec{E}_0$. Using similar approximations for the enhancement factors (eqs. 5 and 6), we could successfully describe several recent experiments.[8,18]

## 2. Transfer times and correlation functions

To describe exciton transfer between SNPs and energy dissipation due to MNPs we should compute the transfer energy rates $\gamma_{metal,i\beta}$ and $\gamma_{Forster,1\beta \to 2\beta'}$. The plasmons and excitons in our system interact via Coulomb fields and we are going to explore this interaction. The total transfer rates for an exciton $|1\beta\rangle$ are given by the Fermi's golden rule:

$$\gamma_{1\beta} = \gamma_{metal,1\beta} + \gamma_{Forster,1\beta} = \frac{2\pi}{\hbar} \sum_{\chi} |\langle \chi | \hat{U}_{Coul} | 1\beta \rangle|^2 \delta(E_1 - E_\chi), \qquad (7)$$

where $E_1 = \hbar\omega_1$ is the exciton energy of the states $|1\beta\rangle$ in SNP1, $\hat{U}_{Coul}$ is the inter-NP Coulomb interaction, and $|\chi\rangle$ are the collective states of the system. These states $|\chi\rangle$ include plasmons in MNPs and excitons in SNP2 and have energies $E_\chi$. In the states $|\chi\rangle$, there is no the exciton in SNP1 since the SNP1 exciton is assumed to be transferred to the other NPs. Eq. 7 describes two types of processes: energy transfer to the metal and FT. The FT rate is composed of transitions from the low-energy exciton state $|1\beta\rangle$ in SNP1 to the excited states of exciton in SNP2 $|exc_1\rangle$: $\gamma_{Forster,1\beta} = \sum_{exc_2} \gamma_{Forster,1\beta \to exc_2}$.

For the SNP2, we have a similar equation

$$\gamma_{2\beta} = \gamma_{metal,2\beta} = \frac{2\pi}{\hbar} \sum_{\chi} |\langle \chi | \hat{U}_{Coul} | 2\beta \rangle|^2 \delta(E_2 - E_\chi), \qquad (8)$$

where $E_2 = \hbar\omega_2$ are the exciton energy of the states $|2\beta\rangle$ in SNP2. Since $E_2 < E_1$ the rate $\gamma_{metal,2\beta}$ does not include the FT process. Again, the states $|\chi\rangle$ include plasmons in MNPs and have no exciton in SNP2. The above approach is based on the perturbation theory. This approach assumes that the excitons are well defined quasi-particles and the exciton-plasmon interaction is relatively weak. In other words, $\hbar\gamma_{i\beta} << E_i$.

Now we express the rates (7 and 8) through correlation functions. The states involved in eq. 7 can be written as $|1\beta\rangle = |0;1\beta\rangle$ and $|\chi\rangle = |f;0\rangle$, where the second index describes the SNP1 states and the first belongs to the rest (MNPs and SNP2). In other words, $\beta$-exciton $|0;1\beta\rangle$ turns into another excitation $|f;0\rangle$. Then, we can write the Coulomb matrix element as $\langle\chi|\hat{U}_{Coul}|1\beta\rangle = \langle f;0|\hat{U}_{Coul}|0;1\beta\rangle = \langle f|\hat{V}_{Coul}|0\rangle$, where

$$\hat{V}_{Coul} = \sum_i \langle 0|\sum_k \frac{e^2}{|\vec{r}_k - \vec{r}_i|}|1\beta\rangle;$$

here, the index $i$ numbers all electrons in the system excluding the electrons involved in the SNP1 excitons and the index $k$ belongs to the electrons contributing to the SNP1 excitons. In typical SNPS, there are four electrons involved in the low-energy excitons. The integral in the matrix element $\langle 0|\frac{e^2}{|\vec{r}_{k'} - \vec{r}_i|}|1\beta\rangle$ should be taken over the coordinates $\vec{r}_k$ (k=1,2,3,4) and $|1\beta\rangle$ is the exciton wave function. To evaluate this integral, we can use the method used in Ref.[20]. The integral should be replaced by a sum of integrals over unit cells inside a SNP and the function $1/|\vec{r}_k - \vec{r}_i|$ should be expanded within a single unit cell. As a result, we obtain:

$$\langle 0|\sum_k \frac{e^2}{|\vec{r}_k - \vec{r}_i|}|\beta\rangle = e^2 d_{exc} \frac{\vec{e}_\beta \cdot (\vec{r}_j - \vec{r}_{SNP1})}{(\vec{r}_j - \vec{r}_{SNP1})^3}$$ for $\vec{r}_j$ is outside of the SNP1. This potential corresponds to a local dipole density $\vec{d}_\beta(\vec{r}) = \vec{e}_\beta \cdot d_{exc} \Psi^2(\vec{r} - \vec{r}_{SNP1})$; the corresponding charge density is $\rho_\beta(\vec{r}) = -d_{exc}\vec{e}_\beta \cdot \partial\Psi^2(\vec{r} - \vec{r}_{SNP1})/\partial\vec{r}$. The Coulomb operator takes

now the form: $\hat{V}_{Coul} = \sum_i V_0(\vec{r}_j - \vec{r}_{SNP1}) = e^2 d_{exc} \sum_i \frac{\vec{e}_\beta \cdot (\vec{r}_j - \vec{r}_{SNP1})}{(r_j - \vec{r}_{SNP1})^3}$, where the potential energy $V_0(\vec{r}) = e^2 d_{exc} \frac{(\vec{e}_\beta \cdot \vec{r})}{r^3}$ for $\vec{r}$ outside the SNP1. Inside SNP1, $V_0(\vec{r})$ is given by the Poisson equation: $\nabla^2 V_0(r) = 4\pi e^2 d_{exc} \vec{e}_\beta \cdot \partial \Psi^2(\vec{r})/\partial \vec{r}$. Using the standard methods[21], we can rewrite eq. 7 in terms of the correlation function:

$$\gamma_{1\beta} = \frac{1}{\hbar^2} \int_{-\infty}^{\infty} \langle 0|\hat{V}^*_{Coul}(t) \cdot \hat{V}_{Coul}|0\rangle e^{iE_1 t/\hbar} dt, \qquad (9)$$

where $\hat{V}^*_{Coul}(t) = e^{i\hat{H}t/\hbar} \hat{V}^*_{Coul} e^{-i\hat{H}t/\hbar}$. Applying the fluctuation-dissipation theorem[21], we obtain

$$\gamma_{1\beta} = \frac{2}{\hbar} \frac{1}{(e^{-\frac{E_1}{k_B T}} - 1)} \operatorname{Im} F(E_1), \qquad (10)$$

where the response function $F(E_1)$ is given by

$$F(E_1) = \frac{-i}{\hbar} \int_0^\infty dt \langle [V^*_{Coul}(t), V_{Coul}] \rangle e^{iE_1 t/\hbar - t/\tau}, \qquad (11).$$

where $\tau \to \infty$. According to the general theory, the response function should be found as a response to the "driving" potential $\Phi_{driving}(\vec{r},t) = \Phi_0(\vec{r}) \cdot e^{-i\omega_1 t}$:

$$F(E_1) \cdot e^{-i\omega_1 t} = \int d^3 r \cdot \Phi_0^*(r) \cdot \rho(r,t) = \int d^3 r \cdot \Phi_0^*(r) \cdot \rho(r) e^{-i\omega_1 t} = \int d^3 r \cdot V_0^*(r) \cdot n(r) e^{-i\omega_1 t}$$

$$(12)$$

where $\rho(r,t) = \rho(r)e^{-i\omega_1 t} = e \cdot n(r)e^{-i\omega_1 t}$ is the non-equilibrium charge density in the system of NPs in the presence of the "driving" potential $\Phi_0(r) \cdot e^{-i\omega_1 t}$, and $\Phi_0(\vec{r}) = V_0(\vec{r} - \vec{R}_{SNP1})/e = ed_{exc} \dfrac{(\vec{e}_\beta \cdot [\vec{r} - \vec{R}_{SNP1}])}{|\vec{r} - \vec{R}_{SNP1}|^3}$ is the dipole potential created by the exciton in SNP1 and the center of SNP1 correspond to $\vec{r} = \vec{R}_{SNP1}$. The factor $(e^{-\frac{E_1}{k_B T}} - 1) \approx -1$, since at room temperature $E_1/k_B T \gg 1$.

Mathematically, in order to obtain the response function $F(E_1)$, we have to compute the function $\rho(r)$ and then take an integral over $\vec{r}$. The integral (12) and the transfer rate (10) can be rewritten as:

$$\operatorname{Im} F(\omega_1) = \operatorname{Im} \int d^3 r \cdot \rho(r) \Phi_0^*(r) = \operatorname{Im} \dfrac{\int d^3 r \cdot \vec{j}(r) \vec{E}_{tot}^*(r)}{i\omega_1},$$

$$\gamma_{1\beta} = \gamma_{metal,1\beta} + \gamma_{Forster,1\beta} = -\dfrac{2}{\hbar} \operatorname{Im} F(\omega_1) = \dfrac{2}{\hbar \omega_1} \operatorname{Im} i \int d^3 r \cdot \vec{j}(r) \vec{E}_{tot}^*(r), \quad (13)$$

where $\vec{j}(r)e^{-i\omega_1 t}$ is the electric current density and $\vec{E}_{tot}(r)e^{-i\omega_1 t}$ is the total field; $\vec{E}_{tot}(r) = -\vec{\nabla}\Phi_0 + \vec{E}_{ind} = -\vec{\nabla}\Phi_{tot}$, where $\vec{E}_{ind}$ is the field of induced charges. To obtain eq. 13, we used two equalities: $-i\omega_1 \rho(r) + \vec{\nabla}\vec{j}(r) = 0$ and $\operatorname{Im}[\int d^3 r \cdot \rho(r)\Phi_{ind}^*(r)] \equiv 0$.

The right-hand side of Eq. 13 is very convenient for calculations since it has a form of an integral of the current density over the space. In fact, the product $\vec{j}(r)\vec{E}_{tot}^*(r)$ is the local heat dissipation (local Joule heat). We also can separate the contribution to the current due to bound charges, plasmons, and excitons. The total potential is given by the Poisson equation:

$$\vec{\nabla}\varepsilon(\vec{r},\omega_1)\vec{\nabla}\Phi_{tot}(\vec{r}) = 4\pi e d_{exc}\vec{e}_\beta \cdot \partial \Psi^2(\vec{r}-\vec{R}_{SNP1})/\partial\vec{r}, \quad (14)$$

where $\varepsilon_1(\vec{r},\omega_1)$ is the local dielectric constant. Outside NPs $\varepsilon(\vec{r},\omega_1) = \varepsilon_0$, inside metal MNPs $\varepsilon(\vec{r},\omega_1) = \varepsilon_m(\omega)$, and inside SNPs $\varepsilon(\vec{r},\omega_1) = \varepsilon_s$. For a single SNP, $\Phi_{tot}(\vec{r}) = ed_{exc}\dfrac{(\vec{e}_\beta \cdot \Delta\vec{r})}{\varepsilon_{eff}\Delta r^3}$ for $|\Delta\vec{r}| > R_{SNP1}$, where $\Delta\vec{r} = \vec{r} - \vec{r}_{SNP1}$, $\varepsilon_{eff} = (\varepsilon_s + 2\varepsilon_0)/3$, and $R_{SNP1}$ and $\vec{r}_{SNP1}$ are the SNP1 radius and position, respectively. The above potential does not include a contribution from the charges induced by the excitons in SNP2. The reason is that the exciton dipole moment is relatively small and its contribution to the induced potential can be neglected compared with dipoles induced by MNPs and bound charges at SNP surfaces. However, we should include the electric currents due to the exciton in SNP2 in the next step. The electric current can be split into two terms: $\vec{j}(r) = \vec{j}_{elect}(r) + \vec{j}_{exc}(r)$. The currents due to bound charges and plasmons are given by: $\vec{j}_{elect}(r) = -i\omega_1 \dfrac{\varepsilon(\vec{r},\omega_1)-1}{4\pi}\vec{E}_{tot}(\vec{r})$. The excitons inside SNP2 create an additional current: $\vec{j}_{exc}(r)$. Therefore, the total rate can be split into two parts:

$$\operatorname{Im} F(\omega_1) = \operatorname{Im}\dfrac{1}{i\omega_1}\int d^3r \cdot \vec{j}(r)\vec{E}^*_{tot}(r) = -\operatorname{Im}\int_{metal} d^3r \dfrac{\varepsilon_m(\omega_1)}{4\pi}\vec{E}_{tot}\vec{E}^*_{tot} + \operatorname{Im}\int_{SNP2} d^3r \cdot \dfrac{\vec{j}_{exc}\vec{E}^*_{tot}}{i\omega_1},$$

(14)

where the first and second terms describe energy transfer processes $SNP1 \rightarrow MNPs$ and $SNP1 \rightarrow SNP2$, respectively. To derive eq. 14, we assumed that $\text{Im}\varepsilon_0 = \text{Im}\varepsilon_s = 0$ and $\text{Im}\varepsilon_m \neq 0$. The physical meaning of the function $\vec{j}(r)\vec{E}_{tot}^*(r)$ inside the integrals in eq. 13 and 14 is the local Joule heat. For the rate of transfer to the MNPs, we obtain

$$\gamma_{metal,1\beta} = -\frac{2}{\hbar}\text{Im}\int_{all\,NPs}d^3r \cdot \rho_{elect}(r)\Phi_0^*(r) = \frac{2}{\hbar}\text{Im}\int_{metal}d^3r\frac{\varepsilon_m(\omega_1)}{4\pi}\vec{E}_{tot}\vec{E}_{tot}^*, \quad (15)$$

where $\rho_{elect}(r)$ is the surface charges induced by the current $\vec{j}_{elect}(r)$. We should stress that these charges appear also on the surface of both SNPs. For the FT rate, we should evaluate the second integral in eq. 14:

$$\gamma_{Forster,1\beta} = \sum_{exc_2}\gamma_{Forster,1\beta \rightarrow exc_2} = -\frac{2}{\hbar}\text{Im}\int_{SNP2}d^3r \cdot \frac{\vec{j}_{exc}\vec{E}_{tot}^*}{i\omega_1}.$$

To calculate the current due to the excited states of exciton $|exc_2\rangle$ in SNP2, we employ the equation of motion of the density matrix (see the Appendix 1). After the standard calculations[22], we obtain:

$$\gamma_{Forster,1\beta}(E_1) = \frac{2\pi}{\hbar}e^2 d_{exc}^2 \sum_{exc_2}\left|\int_{SNP2}d^3r \cdot \Psi_{exc_2}^2(r)\left(\vec{E}_{tot,\beta} \cdot \vec{e}_{exc_2}\right)\right|^2 \frac{\Gamma_{exc_2}}{\pi[\Gamma_{exc_2}^2 + (E_1 - E_{exc_2})^2]}.$$

(16)

Here the state $|exc_2\rangle$ has an electron envelope function $\Psi_{exc_2}(r)$ and a interband dipole moment parallel to the unit vector $\vec{e}_{exc_2}$ ($\vec{e}_{exc_2} = \vec{x}, \vec{y}, \vec{z}$); the state $|exc_2\rangle$ is described by two quantum indexes ($\vec{e}_{exc_2}, k$), where $k$ is the number of energy level and $\vec{e}_{exc_2}$ determines the dipole moment direction. The parameters $\Gamma_{exc_2}$ and $E_{exc_2}$ are the off-diagonal broadening and energy of the excited state $|exc_2\rangle$ of SNP2. Then $\Psi_{exc_2}(r) = \Psi_k(r)$ and $E_{exc_2} = E_k$. Since three vectors $\vec{e}_{exc_2} = \vec{x}, \vec{y}, \vec{z}$ are orthogonal:

$$\sum_{exc_2} \left| \int_{SNP2} d^3r \cdot \Psi_{exc_2}^2(r) \left( \vec{E}_{tot,\beta} \cdot \vec{e}_{exc_2} \right) \right|^2 = \sum_k \left| \int_{SNP2} d^3r \cdot \Psi_k^2(r) \vec{E}_{tot,\beta} \right|^2.$$

We now can make one more step further and include the broadening of the initial exciton state $|1\beta\rangle$. After simple algebra (see Appendix 2), we arrive to

$$\tilde{\gamma}_{Forster,1\beta} = \frac{2\pi}{\hbar} e^2 d_{exc}^2 \sum_k \left| \int_{SNP2} d^3r \cdot \Psi_i^2(r) \vec{E}_{tot,\beta} \right|^2 J(E_1 - E_k),$$

$$J(E_k - E_j) = \int \frac{\Gamma_{exc_2} \Gamma_1}{\pi^2 [\Gamma_{exc_2}^2 + (E' - E_k)^2][\Gamma_1^2 + (E' - E_j)^2]} dE', \qquad (17)$$

where $J(E_1 - E_k)$ is the spectra overlap integral.[1] Now eq. 17 for FT incorporates both off-diagonal broadenings, $\Gamma_{exc_2}$ and $\Gamma_1$. It is not necessary to involve the broadening $\Gamma_1$ for modification of the transfer rate to the metal (eq. 15). The reason is that the plasmon has typically a very broad resonance (about *100 meV*) whereas $\Gamma_1$ is typically about ten *meV*.

For SNP2, we have only transfer to the metal:

$$\gamma_{metal,2\beta} = -\frac{2}{\hbar} \text{Im} \int_{all\ NPs} d^3r \cdot \rho_c(r)\Phi_0^*(r) = \frac{2}{\hbar} \text{Im} \int_{metal} d^3r \frac{\varepsilon_m(\omega_1)}{4\pi} \vec{E}_{tot}\vec{E}_{tot}^*, \quad (18)$$

where the non-equilibrium density and electric fields should be found as response to the "driving" potential $\Phi_0(\vec{r}) = ed_{exc}\frac{(\vec{e}_\beta \cdot [\vec{r} - \vec{r}_{SNP2}])}{|\vec{r} - \vec{r}_{SNP2}|^3}$ ($|\vec{r} - \vec{r}_{SNP2}| > R_{SNP2}$); this potential originates from the exciton in SNP2.

Eqs. 15, 17, and 18 are very convenient for calculation of the transfer rates. In the next section, we will calculate analytically these rates for the dipole regime of interaction.

## 3. Förster transfer rate and efficiency

To describe the effect of MNPs on FT, we introduce the following quantities:

$$f_\beta = \frac{\gamma_{Forster,1\beta}}{\gamma_{Forster,1\beta}^0}, \quad f = \frac{\gamma_{Forster}}{\gamma_{Forster}^0}, \quad e = \frac{\gamma_{Forster} n_1}{I_{1,tot}} = \frac{\gamma_{Forster}}{\gamma_1}. \quad (19)$$

The first two parameters are the relative Förster rates; here $\gamma_{Forster,1\beta}^0$ and $\gamma_{Forster}^0$ denote the FT rates in the absence of MNPs. The parameters $f_\beta$ and $f$ can be regarded as coefficients of plasmon enhancement of FT. The second is the efficiency of FT that is defined as a ratio: $\frac{\text{number of excitons transferred from SNP1 to SNP2}}{\text{number of excitons generated inside SNP1}} = \frac{\gamma_{Forster} n_1}{I_{1,tot}}$. In the absence of energy transfer, $e = 0$. In the presence of strong energy dissipation due the metal, the parameter $e \to 0$.

## 4. Dipole limit

### 4.1 Energy transfer to metal nanoparticles

Fig. 2a shows the geometry. We now use eq. 15 and apply the dipole condition: $d \gg R_{SNP1}, R_{MNP}$, where $d$ is the inter-NP distance. The electric field generated by the exciton $|1\beta\rangle$ outside SNP1 along the z-axis is: $\vec{E}_{exc} = b_\beta e d_{exc} \vec{e}_\beta / (\varepsilon_{eff} d^3)$, where $b_x = b_y = -1$ and $b_z = 2$, $\varepsilon_{eff} = (\varepsilon_s + 2\varepsilon_0)/3$; $\vec{e}_\beta$ are the unit exciton-polarization vectors as before. The field $\vec{E}_{exc}$ is partially screened by the background dielectric constant of SNP. Then, the resultant field inside the MNP becomes "screened" one more time[19]:

$$\vec{E}_{tot} = \frac{b_\beta e d_{exc} \vec{e}_\beta}{\varepsilon_{eff} d^3} \frac{3 \cdot \varepsilon_0}{\varepsilon_m(\omega) + 2 \cdot \varepsilon_0}.$$

Assuming that $\vec{E}_{tot}$ does not vary much over the MNP volume, we obtain from eq. 15 (right-hand side form) for the energy transfer rate:

$$\gamma_{metal,1\beta} = -\frac{2 \cdot b_\beta^2}{\hbar} \frac{e^2 d_{exc}^2}{d^6 \varepsilon_{eff}^2} \frac{3 \cdot \varepsilon_0^2 \cdot R_{MNP}^3}{|\varepsilon_m(\omega) + 2 \cdot \varepsilon_0|^2} \operatorname{Im} \varepsilon_m(\omega). \qquad (20)$$

Using this formula, it was possible to describe experimental data of several groups.[18] The $1/d^6$-dependence corresponds to the FT theory. In the other limit $\Delta << R_{MNP}$ and $\Delta >> R_{SNP1}$, $\gamma_{metal,1\beta} \sim d^{-3}$ (see ref.[23]). Here $\Delta$ is the surface-to-surface distance (Fig. 2a).

**4.2 The role of bound charges in transfer processes**

It is seen from eq. 15 that the rate $\gamma_{metal,1\beta}$ can be written either through the charge density or through the current density. It is interesting that the correct expression for the transfer rate, that is written through the charge density, should include two terms:

$$\gamma_{metal,1\beta} = -\frac{2}{\hbar} \text{Im} \left[ \int_{SNP1} d^3r \cdot \rho_{elect}(r) \Phi_0^*(r) + \int_{MNP} d^3r \cdot \rho_{elect}(r) \Phi_0^*(r) \right]. \quad (21)$$

The first term is due to the image charges on the surface of SNP1. Even though the constants $\varepsilon_0$ and $\varepsilon_s$ have not an imaginary part ($\text{Im}\,\varepsilon_{0,s} = 0$), the first term contributes to the dissipation. Mathematically, the reason is that these charges are partially induced by the dipole moment of the MNP and this dipole moment includes $\varepsilon_m(\omega)$ with a nonzero imaginary part ($\text{Im}\,\varepsilon_m(\omega) \neq 0$). Simultaneously, the current form for the transfer rate $\gamma_{metal,1\beta}$ includes only an integral over MNP, i.e.

$$\gamma_{metal,1\beta} = -\frac{2}{\hbar} \text{Im} \int_{MNP} d^3r \frac{\varepsilon_m(\omega_1)}{4\pi} \vec{E}_{tot} \vec{E}_{tot}^*. \quad (22)$$

Since the function inside this integral is the local Joule-heat dissipation rate, this form indicates that the dissipation process appears inside the MNP, whereas the formula (21) is another mathematical representation for the integral (22).

The role of the bound surface charges on the SNP1 surface is important numerically since, for most matrixes, $\varepsilon_0$ is essentially large than unit. For example, $\varepsilon_0 \approx 1.8$ and 2.3 for water and polymers at the optical energies $E_1 \sim 2\,eV$. We can see from eq. 20 that the metal transfer rate strongly depends on $\varepsilon_0$. In the case of matrix with $\varepsilon_0 > 1$, the bound charges originate both from NPs and matrix. Fig. 2d illustrates this situation. If we formally introduce a thin vacuum layer between NPs and matrix, we can see that bound charges can be accumulated on both interfaces. Since our general formalism treats the surface changes consistently, the general formulas (15, 18) incorporate the surface bound charges originating from both NPs and matrix. We also note that the operator $\hat{V}_{Coul}$ in eq. 9 includes summation over electrons of the matrix (e.g. water).

**4.3 Förster transfer**

Now we calculate the FT rate from eq. 17 in the absence of MNPs (Fig. 2b). The electric field induced by the exciton $|1\beta\rangle$ inside SNP2 is $\vec{E}_{tot} = \dfrac{b_\beta e d_{exc}\vec{e}_\beta}{\varepsilon_{eff} d^3} \dfrac{3\cdot\varepsilon_0}{\varepsilon_s + 2\cdot\varepsilon_0}$. The sum in eq. 17 should be taken over three bright excitons with $\vec{e}_{exc_2} = \vec{x}, \vec{y}, \vec{z}$ (Fig. 2b). The FT rate is

$$\tilde{\gamma}_{Forster,1\beta} = \frac{2\pi}{\hbar} \frac{e^4 d_{exc}^4 b_\beta^2}{d^6} \frac{\varepsilon_0^2}{\varepsilon_{eff}^4} J(E_1), \qquad (23)$$

where

$$J(E_1) = \sum_{exc_2} J(E_1 - E_{exc_2})$$

is the overlap integral. Equation (23) corresponds to the Förster theory[1] and also includes the screening effect of bound charges ($\varepsilon_0^2 / \varepsilon_{eff}^4$). Again, we see the importance of surface charges.

**4.4 Plasmon-assisted Förster transfer**

Now we focus on the FT rate in the presence of MNP (Fig. 2c). In the dipole limit, the total electric field inside SNP2 created by the exciton $|1\beta\rangle$ is a sum of two contributions coming from SNP1 and MNP:

$$\vec{E}_{tot,\beta} = \frac{\varepsilon_0}{\varepsilon_{eff}} \frac{ed_{exc}}{\varepsilon_{eff}} \frac{3(\vec{e}_\beta \cdot \vec{r}_{12})\vec{r}_{12} - r_{12}^2 \vec{e}_\beta}{r_{12}^5} + \frac{\varepsilon_0}{\varepsilon_{eff}} \frac{3(\vec{D}_{MNP} \cdot \vec{r}_{m2})\vec{r}_{m2} - r_{m2}^2 \vec{D}_{MNP}}{r_{m2}^5},$$

where

$$\vec{D}_{MNP} = \beta_{MNP}(\omega) \cdot \vec{E}_{SNP1}^0$$

is the dipole moment induced in MNP by the SNP1 exciton, $\beta_{MNP}(\omega) = R^3 \frac{\varepsilon_m - \varepsilon_w}{(2\varepsilon_w + \varepsilon_m)}$,

$$\vec{E}_{SNP1}^0 = \frac{ed_{exc}}{\varepsilon_{eff}} \frac{3(\vec{e}_\beta \cdot \vec{r}_{1m})\vec{r}_{1m} - r_{1m}^2 \vec{e}_\beta}{r_{1m}^5}, \quad \vec{r}_{1m} = \vec{r}_m - \vec{r}_1, \quad \vec{r}_{12} = \vec{r}_2 - \vec{r}_1, \text{ and } \vec{r}_{m2} = \vec{r}_2 - \vec{r}_m; \quad \vec{r}_1, \vec{r}_2,$$

and $\vec{r}_m$ are the coordinates of SNP1, SNP2, and MNP, correspondingly. The bright excited states in SNP2 have the polarizations $\vec{e}_{exc_2} = \vec{x}, \vec{y}, \vec{z}$. Eq. 17 takes a form:

$$\tilde{\gamma}_{Forster,1\beta} = \frac{2\pi}{\hbar} e^2 d_{exc}^2 J(E_1) \cdot \left| \vec{E}_{tot,\beta} \cdot \right|^2, \qquad (24)$$

The terms in this sum are the rates $\tilde{\gamma}_{1\beta \to exc_2}$ for the processes $|1\beta\rangle \to |exc_2\rangle$. The field enhancement factor for FT

$$f_\beta = \frac{\tilde{\gamma}_{Forster,1\beta}}{\tilde{\gamma}^0_{Forster,1\beta}} = \frac{\left|\vec{E}_{tot,\beta} \cdot \right|^2}{\left|\vec{E}_{tot,\beta,nometal}\right|^2} =$$

$$= \frac{\left| ed_{exc} \frac{\varepsilon_0}{\varepsilon_{eff}^2} \frac{3(\vec{e}_\beta \cdot \vec{r}_{12})\vec{r}_{12} - r_{12}^2 \vec{e}_\beta}{r_{12}^5} + \frac{\varepsilon_0}{\varepsilon_{eff}} \frac{3(\vec{D}_{MNP} \cdot \vec{r}_{m2})\vec{r}_{m2} - r_{m2}^2 \vec{D}_{MNP}}{r_{m2}^5} \right|^2}{\left| ed_{exc} \frac{\varepsilon_0}{\varepsilon_{eff}^2} \frac{3(\vec{e}_\beta \cdot \vec{r}_{12})\vec{r}_{12} - r_{12}^2 \vec{e}_\beta}{r_{12}^5} \right|^2}.$$

(25)

In the end of this section, we give simple equations for the case $\varphi = 180^0$ (Fig. 2c):

$$\tilde{\gamma}_{Forster,1x} = \tilde{\gamma}_{Forster,1y} = \frac{2\pi}{\hbar} e^4 d_{exc}^4 \left(\frac{\varepsilon_0}{\varepsilon_{eff}^2}\right)^2 \left| -\frac{1}{d^3} + \frac{\beta_{mnp}(\omega)}{d_1^3 d_2^3} \right|^2 J(E_1),$$

$$\tilde{\gamma}_{Forster,1z} = \frac{2\pi}{\hbar} e^4 d_{exc}^4 \left(\frac{\varepsilon_0}{\varepsilon_{eff}^2}\right)^2 \left| \frac{2}{d^3} + \frac{4 \cdot \beta_{mnp}(\omega)}{d_1^3 d_2^3} \right|^2 J(E_1). \qquad (26)$$

Note that, for the geometry of Fig. 2c, the vector $\vec{n}$ is parallel to $\vec{z}$. From eq. 23, it is easy to see that the plasmon effect on the FT rate for the SNP1 exciton $\beta = z$ should be the largest, because of the factor $4^2$. The reason is that the optical dipole for the state $\beta = z$ is perpendicular to the MNP surface.

**4.5 Numerical results for plasmon-assisted Förster transfer in the dipole limit**

The NP position vectors in a rather complicated eq. 22 can be expressed, for convenience, through the inter-NP distances $d_1$, $d_2$, and $d$, and the angle $\varphi$ (Fig. 2c). Figs. 3, 4, and 5 show the results for Ag MNP. For the Ag dielectric constant $\varepsilon_m(\omega)$, we use empirical bulk values from ref.[24]. For the other dielectric constants, we took $\varepsilon_0 = 1.8$ (water) and $\varepsilon_s = 7.2$ (CdTe). The geometrical parameters were somewhat similar to the experiments[4,18]: $d_1 = d_2 = 10nm$ and $R_{NMP} = 4nm$. The optical dipole moment can be estimated from the typical radiative lifetime. For CdTe and CdSe SNPs, a radiative lifetime measured in time-resolved PL studies at room temperature $1/\gamma_{rad}^0 \approx 10ns$. Since there are three bright and five dark excitons, $\gamma_{rad}^0 = (3/8)\gamma_{rad,exc}^0$, where $\gamma_{rad,exc}^0 = \gamma_{rad,\beta}^0$ is the radiative rate of a bright exciton ($\beta = x, y, z$). From the quantum optics[22], we know that

$$\gamma_{rad,exc}^0 = \frac{8\pi\sqrt{\varepsilon_0}\omega_{exc}^3 e^2 d_{exc}^2}{3(\varepsilon_{eff}/\varepsilon_0)^2 h \cdot c^3}. \tag{27}$$

Using the above value for $\gamma_{rad}^0$, we obtain an estimate: $d_{exc} \approx 4.6 A$. Another important parameter of the problem is the overlap integral, $J$. From experimental studies,[4] we can estimate the typical FT rate as $\gamma_{Forster} = \sum_{\alpha'\alpha''}\gamma_{Forster,1\alpha \to 2\alpha''}/8 \sim 1/(10ns)$ for $d \sim 10nm$.

From these numbers, we obtain an estimate: $J \sim 0.008 \, meV^{-1}$. The FT parameters ($f_\beta$ and $\gamma_{Forster,1\beta}$) are calculated using eqs. 24 and 25. The angle $\varphi$ was taken as $60^0$ and $180^0$ for Fig. 3a and b, respectively. We can see significant enhancement of FT for the wavelengths close the plasmon resonance. Orientation of exciton dipole relative to the MNP surface plays an important role. Strongest enhancement can be achieved if the exciton dipole is perpendicular to the surface of MNP; this is because the MNP plasmon resonance mostly enhances electric fields perpendicular to the MNP surface.[18] Fig. 3b also includes numerical results (see Sec. 5.1). We can see that the analytical results obtained within the dipole approximation provide us with reasonable estimates for the NP complex with $d_{1(2)}/R_{MNP} = 0.4$ (error is about 10%).

Another important geometrical parameter is the angle $\varphi$. Figs. 4 and 5 show the spatial maps of the FT coefficient. As a two-directional variable, we use the SNP2 position, $\vec{r}_{SNP2}$. The position of SNP1 is fixed: $\vec{r}_{SNP1} = (0, 0, 10 \, nm)$. The geometry with $\varphi = 180^0$ has clear advantage from the point of view of FT rate enhancement. Simultaneously, the geometry with $\varphi = 180^0$ has obvious disadvantage: the inter-SNP distance for $\varphi = 180^0$ is relatively long and the absolute value of the FT rate can be relatively small; therefore, the FT effect would be less visible in optical spectra.

Calculated FT rate is shown in Figs. 6 and 7. One can see that the plasmon resonance effect on the FT process. Plasmon-enhancement of FT can be seen for the $z$ − exciton in the geometry of $\varphi = 60^0$ in the vicinity of the plasmon resonance in Ag NP (below $400 \, nm$) and in the geometry of $\varphi = 180^0$ for $\lambda > 400 \, nm$. We also can see the effects of screening and suppression of FT process. In particular, $z$-exciton transfer becomes strongly enhanced for the geometry $\varphi = 180^0$ (Fig. 7, middle). Such process can be called as plasmon-assisted *long-distance Förster transfer*.

**4.4 Diagram representation for the plasmon-enhanced FT**

We now derive eq. 26 using the diagram method. We start from the expression (7) and apply the standard perturbation theory method. To the second order of the perturbation theory, the amplitude of FT is given by[25]:

$$\langle \chi | \hat{U}_{Coul} | 1\beta \rangle = \langle 0; exc_2; 0_{pl} | \hat{U}_{SNP1-SNP2} | 1\beta; 0_{exc_2}; 0_{pl} \rangle +$$
$$\sum_{pl'} \frac{\langle 0; exc_2; 0_{pl} | \hat{U}_{SNP1-MNP} | 0_{exc_1}; 0_{exc_2}; pl' \rangle \langle 0_{exc_1}; 0_{exc_2}; pl' | \hat{U}_{SNP2-MNP} | 1\beta; 0_{exc_2}; 0_{pl} \rangle}{E_1 - E_{pl'} + i\hbar/\tau}, \quad (28)$$

where the wave function $|1\beta; 0_{exc_2}; 0_{pl}\rangle$ describes the state with one exciton $|1\beta\rangle$ in SNP1, and no plasmons and excitons in MNP and SNP2; similarly, the wave function $|0_{exc_1}; 0_{exc_2}; pl'\rangle$ denotes the state with one plasmon and no excitons. The total Coulomb interaction operator should be written as: $\hat{U}_{Coul} = \hat{U}_{SNP1-SNP2} + \hat{U}_{SNP1-MNP} + \hat{U}_{SNP2-MNP}$, where the terms describe the dipole-dipole interactions between three NPs. For example, $\hat{U}_{SNP1-SNP2} = \frac{\varepsilon_0 e^2}{\varepsilon_{eff}^2} \sum_{1,2} \frac{\delta \vec{r}_1 \cdot \delta \vec{r}_2 - 3(\delta \vec{r}_1 \cdot \vec{n})(\delta \vec{r}_2 \cdot \vec{n})}{d^3}$, where $\delta \vec{r}_1$ and $\delta \vec{r}_2$ are electron coordinates related to SNP1 and SNP2, respectively; the sum should be taken over all electrons taking part in inter-band transitions in SNPs. The unit vector $\vec{n}$ "connects" the SNPs (Fig. 2c). For the SNP-MNP interaction, we have $\hat{U}_{SNP1-SNP2} = \frac{e^2}{\varepsilon_{eff}} \sum_{1,m} \frac{\delta\vec{r}_1 \cdot \delta\vec{r}_m - 3(\delta\vec{r}_1 \cdot \vec{n})(\delta\vec{r}_m \cdot \vec{n})}{d_1^3}$, where $\delta\vec{r}_m$ is the position of an electron inside MNP; again, $\delta\vec{r}_m$ is related to the MNP center. Using the fluctuation-dissipation theorem[21] and the response function of a single spherical MNP, one can obtain the useful equality:

$$\beta_{MNP}(\omega) = -\frac{e^2}{\varepsilon_0} \sum_{pl'} |\langle pl'|x|0_{pl}\rangle|^2 \left\{ \frac{1}{\hbar\omega - E_{pl'} + i\hbar/\tau} - \frac{1}{\hbar\omega - E_{pl'} + i\hbar/\tau} \right\} . \quad (29)$$

In the next step, we use the resonant approximation (i.e. neglecting the second term in the equation 29) and obtain

$$\langle \chi|\hat{U}_{Coul}|1\beta\rangle = \frac{\varepsilon_0}{\varepsilon_{eff}^2} e^2 d_{exc}^2 \left( \frac{1}{d^3} + \frac{b_\beta \beta_{mnp}(\omega)}{d_1^3 d_2^3} \right), \quad (30)$$

where $b_x = b_y = -1$ and $b_z = 2$, as before. This equation reproduces our previous result (eq. 26). The upper two diagrams in Fig. 8a show the processes corresponding to the two terms in eq. 26. The second diagram in Fig. 8a describes the FT assisted by virtual creation of a plasmon in MNP. Using perturbation theory, we also can calculate the FT amplitude in the presence of few MNPs. As an example, we solve now the case of two MNPs ($a$ and $b$) arranged in a line (Fig. 8c). This case has a rather simple geometry; $d_{ab}$ is the inter-MNP distance, and $d_{ia}$ and $d_{ib}$ are the SNP-MNP distances. Including the interaction between MNPs in all orders of the parameter $1/d_{ab}^3$, we arrive to

$$\langle \chi | \hat{U}_{Coul} | 1\beta \rangle =$$

$$= \frac{\varepsilon_0}{\varepsilon_{eff}^2} e^2 d_{exc}^2 \left( \frac{1}{d^3} + \left\{ \frac{b_\beta \beta_a}{d_{1a}^3 d_{2a}^3} + \frac{\varepsilon_0^2}{\varepsilon_{eff}} \frac{b_\beta^2 \cdot \beta_a \cdot \beta_b}{d_{1a}^3 d_{2b}^3 d_{ab}^3} + \frac{b_\beta \beta_b}{d_{1b}^3 d_{2b}^3} + \frac{\varepsilon_0^2}{\varepsilon_{eff}} \frac{b_\beta^2 \cdot \beta_a \cdot \beta_b}{d_{1b}^3 d_{2a}^3 d_{ab}^3} \right\} \frac{1}{1 - \frac{\varepsilon_0^2}{\varepsilon_{eff}^2} \frac{b_\beta^2 \cdot \beta_a \cdot \beta_b}{d_{ab}^6}} \right)$$

where the fuctions $\beta_{a(b)} = \beta_{MNP,a(b)}(\omega)$ are defined for two MNPs (*a* and *b*). The above formula assumes that the interaction between MNPs involves only dipole moments. In the following, we will give numerical results involving all multi-poles.

In the end of this section, we comment on the renormalization of the exciton energy of a single SNP due to the presence of other NPs. Mathematically, the exciton shift due to the presence of other NPs is given by the real part of the response function: $\delta E_{1\beta} \sim \frac{1}{\hbar} \operatorname{Re} F(E_1)$. For the case of SNP-MNP interaction only, this shift is given be the MNP response function: $\delta E_{1\beta} \sim \frac{1}{\hbar} \operatorname{Re} \int_{metal} d^3 r \frac{\varepsilon_m(\omega_1)}{4\pi} \vec{E}_{tot} \vec{E}_{tot}^*$. To calculate the shift using diagrams, one should look at the response function $F(\hbar\omega)$ and sum up an infinite series of diagrams that involve the exciton state; Fig. 8b shows two first relevant diagrams. For SNP-MNP molecule, the exciton shift is a relatively small number (less than *1 meV*) as calculated in ref.[18] This shift of exciton energy can become important at low temperatures whereas, at room temperature, it can be neglected.

## 5. Numerical results

### 5.1 Numerical method

For numerical results, it is necessary to compute reliably electric fields inside a NP complex and the response function and $F(\hbar\omega)$ (eq. 11). Here we employ the multipole expansion method. This method is every efficient and was used by us to calculate optical properties of complexes with many NPs (up to one hundred).[18] To solve the Poisson equation (eq. 14) for the medium with non-uniform dielectric constant $\varepsilon_1(\vec{r},\omega)$, we expand the electrical potential in terms of spherical harmonics: $\varphi_{tot} = \varphi_{external}(t,\vec{r}) + \sum_n \varphi_n$, where $\varphi_{external}(t,\vec{r}) = \varphi_{ext}(\vec{r})e^{-i\omega t}$ and $\varphi_n$ is the potential induced by the charges on the surface of the $n^{th}$-NP. The potentials of single NPs are expanded in terms of spherical harmonics:

$$\varphi_n(\mathbf{r}) = \sum_{l,m} q^n_{l,m} \frac{Y_{l,m}(\theta_n,\phi_n)}{r_n^{l+1}}, \qquad (31)$$

where $Y_{l,m}(\theta,\phi)$ are the spherical harmonics, $-l \leq m \leq l$, and the coordinates $(\theta_n,\phi_n,r_n)$ are related to the coordinate system of the $n^{th}$-NP. The standard boundary conditions are introduced at the surface of each NP and involve the dielectric constants $\varepsilon_0$, $\varepsilon_m$, and $\varepsilon_s$. In our computations, we truncate the system of equations assuming that the coefficients $q^n_{l,m}$ rapidly decrease as a function of $l$. Therefore, we include only $l \leq l_{max}$. Below we will use $l_{max} = 5$ that provides us with a very good precision (the error is a few %).

## 5.2 Silver NPs

Below we will use geometrical parameters similar to those used in Sec.4.5: $d_1 = d_2 = 10 nm$ and $R_{NMP} = 4 nm$. First of all, we show that the dipole approximation provides us with reliable estimates. As an example, Fig. 9a shows the rates of exciton transfer from SNP1 to MNP calculated within the dipole approximation (eq.15) and numerically. We also observed that the dipole approximation gives reliable estimates for the FT rates (Fig. 3b).

First we show the numerical results for the imaginary part of the response function (eq. 7). As an example, we will compute the response function describing the transfer process of $1_z$-exciton:

$$\gamma_{1z} = \gamma_{metal,1z} + \gamma_{Forster,1z} = -\frac{2}{\hbar}\mathrm{Im}\, F(\omega_1) = \frac{2}{\hbar\omega_1}\mathrm{Im}\, i\int d^3r \cdot \vec{j}(r)\vec{E}^*_{tot}(r)$$

(see Fig. 10). This function describes energy transfer from SNP1 to SNP2 and MNP. Mathematically, this function looks like a dissipation rate of a SNP1 exciton due to absorption in SNP2 and MNP. In Fig. 10, the exciton emission wavelength of SNP1 is a variable, whereas the exciton wavelength of SNP2 is fixed ($\lambda_{SNP2} = 600 nm$). The dielectric constants and the overlap integral ($J$) were specified above. We see that $\gamma_{metal,1z} \gg \gamma_{Forster,1z}$. This inequality reflects the fact that MNP has many mobile electrons participating in the energy transfer process. These electrons create a strong plasmon resonance and large dipole. Simultaneously, SNP2 absorbs the SNP1-exciton via the exciton resonance; in our simplified model, the exciton absorption under the resonant condition ($E_1 \approx E_{exc_2}$) involves only 4 valence-band electrons (2 heavy-hole and 2 light-hole states). However, SNPs give the main contribution to photoluminescence process and, therefore, the rate $\gamma_{Forster}$ plays the important role in optical experiments and can be measured.[2,3,4] Emission from metal NPs and their quantum yield are typically small because of fast energy relaxation inside the metal crystal.

Due to fast spin relaxation, the optical spectra of SNPs at room temperature depend on the averaged FT and metal transfer rates:

$$\gamma_{Forster} = \sum_{\beta,\beta'=x,y,x} \gamma_{Forster,1\beta \to 2\beta'}/8, \quad \gamma_{metal,1} = \sum_{\beta} \gamma_{metal,1\beta}/8.$$

These rates can be calculated using the following formulas (see eqs. 16 and 17):

$$\gamma_{metal,1\beta} = -\frac{2}{\hbar} \text{Im} \int_{all\ NPs} d^3r \cdot \rho_{elect}(r) \Phi_0^*(r),$$

$$\gamma_{metal,2\beta} = -\frac{2}{\hbar} \text{Im} \int_{all\ NPs} d^3r \cdot \rho_{elect}(r) \Phi_0^*(r)$$

$$\tilde{\gamma}_{Forster,1\beta} = \frac{2\pi}{\hbar} e^2 d_{exc}^2 \sum_k \left| \int_{SNP2} d^3r \cdot \Psi_k^2(r) \vec{E}_{tot,\beta} \right|^2 J(E_1 - E_k), \quad (32)$$

where the functions $\rho_{elect}(\vec{r})$ and $\vec{E}_{tot,\beta}(\vec{r})$ should be numerically calculated from the Poisson equation (14). These functions are the change density induced on the surfaces of all NPs and the total electric field. The above charges and electric field originate from the oscillating dipole with spatial charge density $\rho_0(r,t) = ed_{exc}(\vec{e}_\beta \cdot \partial \Psi^2(\vec{r} - \vec{R}_{SNP1})/\partial \vec{r}) \cdot e^{-i\omega_1 t}$. The "driving" potential $\Phi_0(r)e^{-i\omega_1 t}$, created by $\rho_0(r,t)$, also enters eq. 32. For the excited states in eq. 32, we use the approximation $\Psi_k^2(r) = 1/V_{SNP2}$. In addition, we assume for simplicity that the overlap integral $J(E_1)$ is a constant: $J(E_1) = J = 0.008\ meV^{-1}$. As it was mentioned above, with this number for the overlap integral, we obtain realistic numbers for inter-SNP FT rates. Figs. 11 and 12 show the averaged FT rate. The plasmon enhancement effect for the geometry $\varphi = 180^0$ is very remarkable. Again, we see plasmon-enhanced, long-range FT.

Now we compute the FT efficiency and the ratio of emission peaks. The corresponding equations involve the bright excitons:

$$e = \frac{\gamma_{Forster}}{\gamma_1} = \frac{\gamma_{Forster}}{\gamma_{rad,1} + \gamma_{metal,1} + \gamma_{Forster}},$$

where the averaged radiative rate is:

$$\gamma_{rad,1} = \gamma_{rad,exc}^0 \sum_\beta P_{1\beta}^a(\omega_1)/8.$$

Calculated efficiency and metal transfer rate are shown in Fig. 13. As expected, the presence of the metal component reduces the FT efficiency since the exciton energy flows to MNP. We also can see the regime of enhanced FT efficiency for $\varphi = 180^0$ and the exciton wavelength $\lambda_{SNP1} > 570nm$. The increase of efficiency in this regime happens because the metal transfer rate becomes small far from the plasmon resonance.

In general, we observe that the FT efficiency is reduced in the vicinity of the plasmon resonance due to strong dissipation.

Now we calculate the PL spectrum of a SNP. The PL peak intensities are given by

$$I_{PL,1} = \frac{\gamma_{rad,1}}{\gamma_1} I_{1,tot}, \quad I_{PL,2} = \frac{\gamma_{rad,2}}{\gamma_2}\left(I_{2,tot} + I_{1,tot}\frac{\gamma_{Forster}}{\gamma_1}\right).$$

We calculated PL spectra as a sum of two Lorentzian peaks: $I_{PL}(\omega) = I_{PL,1}\cdot L(\omega-\omega_{SNP1}) + I_{PL,2}\cdot L(\omega-\omega_{SNP2})$, where $L(\omega) = \Gamma_{PL}/(\Gamma_{PL}^2 + \omega^2)$ and $\Gamma_{PL} = 80\,meV$. Figure 14 shows the results. As before, we used: $1/\gamma_{rad}^0 \approx 10\,ns$ and $J = 0.008\,meV^{-1}$. Also, we assumed that $I_{2,tot} = I_{1,tot}$. Calculated PL spectra are mostly governed by the processes of energy transfer to MNP and it is not easy to see the effect of FT. For the SNP pair $(\lambda_{SNP1}, \lambda_{SNP2}) = (440nm, 500nm)$ and $\varphi = 90^0$, we obtain the following parameters: $\gamma_{Forster}^0 = 0.1\,ns^{-1}$ and $\gamma_{Forster} = 0.78\,ns^{-1}$. The corresponding FT enhancement parameter $f = 7.8$. Simultaneously, the metal transfer rates become large: $\gamma_{metal,1} = 5.2\,ns^{-1}$ and $\gamma_{metal,2} = 1.1\,ns^{-1}$. The asymmetry of peaks in Fig. 14a is mostly due to the large difference between $\gamma_{metal,1}$ and $\gamma_{metal,2}$. For the case of $(\lambda_{SNP1}, \lambda_{SNP2}) = (400nm, 500nm)$ and $\varphi = 180^0$, we obtain the same behavior. The PL peak asymmetry is mostly due to a large difference between metal transfer rates for SNP1 and SNP2. In the case of Fig. 14b, the FT enhancement factor is remarkably large: $f \approx 50$.

To summarize, we see from our calculations that the cw-PL study is not so informative and should be analyzed very carefully using the rate equations incorporating the set of parameters, $\gamma_{Forster}$, $\gamma_{meta,1(2)}$, $\gamma_{rad,1(2)}$, and $I_{1(2),tot}$. In particular, we should comment on the PL peak asymmetry. In many experiments, the PL peak asymmetry can be taken as a signature of FT. But, this is not the case for our situation. Here, the PL peak asymmetry occurs due to exciton energy transfer to MNP. The exciton peak of SNP1 becomes less intensive than that of SNP2 because the energy relaxation rate for SNP1 exciton is larger (see the numbers in Fig. 14). The latter comes from the fact that the SNP1 exciton energy is closer to the plasmon resonance. A time-resolved study of single NP complex may be more productive and we will discuss this opportunity below.

## 5.3 Gold NPs

Another important material system is Au nanocrystals. We now show the results for SNP-MNP complexs with parameters: $d_1 = d_2 = 8nm$, $R_{NMP} = 4nm$, $R_{SMP1} = 1.5nm$, and $R_{SMP2} = 2.5nm$. For the geometry $\varphi = 90^0$, the FT process is suppressed due to the screening effect (Fig. 15a). But, the geometry $\varphi = 180^0$ demonstrates again enhanced, plasmon-assisted FT (Fig. 15b). In Fig. 15b, we also plot the metal energy transfer rate. Again, this rate is larger than $\gamma_{Forster}$. This leads to a small FT efficiency, like in the case of Ag-MNP (Fig. 13).

Like in the case of Ag-MNP, the metal relaxation rates determine the intensity of exciton peaks in cw-PL spectra. AS an example, we show the PL spectrum for $(\lambda_{SNP1}, \lambda_{SNP2}) = (800nm, 900nm)$. Because of fast energy relaxation, the PL emission becomes strongly reduced (inset of Fig. 16). The peak asymmetry comes mostly from the difference in energy relaxation rates, $\gamma_{metal,1}$ and $\gamma_{metal,2}$. The effect of FT is visible in Fig. 16. In experiments, it can be derived from the PL spectrum using the rate equations.

A more promising method to observe plasmon-enhanced FT can be time-resolved PL. We now consider temporal dynamics of excitons and denote the populations of SNPs as be $n_1(t)$ and $n_2(t)$. At $t=0$, $n_1(0) = n_{10}$ and $n_2(0) = n_{20}$ for excitons in SNP1 and SNP2, respectively. The temporal evolution of exciton populations is given by:

$$n_1(t) = n_{10} e^{-\gamma_1 t}, \quad n_2(t) = a \cdot e^{-\gamma_1 t} + b \cdot e^{-\gamma_2 t},$$

where $a = -\gamma_{Forster}/(\gamma_1 - \gamma_2) n_{10}$ and $b = n_{20} + \gamma_{Forster}/(\gamma_1 - \gamma_2) n_{10}$. For simplicity, we assume that $n_{12} = n_{20}$. Fig. 17 shows temporal dynamics. The indication of FT is the presence of two exponential functions in the temporal evolution of PL signal of SNP2. Therefore, we will analyze the ratio $a/b$. For the case $(\lambda_{SNP1}, \lambda_{SNP2}) = (600nm, 660nm)$, $a/b \approx 0.03$. The ratio between pre-exponential amplitudes for the case $(600nm, 660nm)$ is $a/b \approx 0.33$. Therefore, this looks assessable for experimental studies. Fig. 17 shows also the straight line with a slope equal to $d(\ln n_2(t))/dt |_{t=0}$. We can see in Fig. 17 that the functions $\ln n_2(t)$ decrease more rapidly than the straight lines. In this way, one can see the effect of FT. We note that similar dynamics was observed, for example, in ref.[3] for an ensemble of SNPs. We also note that the FT enhancement factors for the data in Fig. 17 are large ($\sim 8, 17$).

## 6. Discussion and conclusions

The FT process is qualitatively different to the process of energy transfer in plasmonic wave guides.[26,27] The main difference is that the FT process can appear without excitation of plasmons. Moreover, direct plasmon excitation leads to dissipation and makes it difficult to observe the FT process. In other words, energy transfer through *virtual* plasmons is the best. In contrast, energy transfer in plasmonic waveguides comes from excitation and propagation of real plasmons; experimentally such transfer was realized in ref.[26]. The ideal situation for plasmon-assisted FT is energy channeling without excitation of real plasmons. This trend can be well seen using an idealized Drude model with very small plasmon damping. We now assume that MNP has the following dielectric constant: $\varepsilon_m(\omega) = \varepsilon_\infty [1 - \Omega_p^2 /(\omega(\omega - i\gamma_p))]$. For the parameters of the model, we take: $\Omega_p = 2.6\,eV$, $\varepsilon_\infty = 12$, and $\gamma_p = 0.005\,eV$. The last parameter is the plasmon damping. Of course, the realistic damping frequencies in metals are much larger. Figure 18 describes the FT process assisted by a metal NP with very small plasmon damping. The ratio $\gamma_{Forter,1z}/\gamma_{metal,1}$ is minimal in the plasmon resonance because of strong energy losses during the FT process. As the exciton energy moves away the plasmon resonance, the ratio $\gamma_{Forter,1z}/\gamma_{metal,1}$ increases and FT becomes more efficient.

Plasmons in optical experiments with metals represent oscillators with strong damping and, therefore, plasmon-assisted FT of excitons should appear together with strong energy relaxation. This tells us that the conditions for experimental observation of plasmon-enhanced FT are peculiar: the enhanced FT process should be studied in the regime of relatively weak dissipation. In this paper, we describe several examples of SNP pairs: $(\lambda_{SNP1}, \lambda_{SNP2}) = (400, 500), (440, 500), (600, 660)$, and $(800 \text{ nm}, 900 \text{ nm})$. CdTe SNP pair can be used for the case 600/660. InGaN can be used for all above pairs as this material has the unique compositional tunability; the band gap of InGaN can be tuned from 0.7 to 3.45 eV.[28] PbSe NPs also offer the exciton energies from about 0.2 to 1.6 eV.[29] PbSe NPs can be used for the pair (800, 900). A disadvantage of PbSe NPs is a relatively high background dielectric constant ($\varepsilon_s = 23$); this will reduce the FT rate. Simultaneously, the dielectric constants of CdTe and InGaN are not very large. For CdTe $\varepsilon_s = 7.2$ and, for InGaN, the dielectric constant is in the range $5.15 < \varepsilon_s < 8.4$.

To summarize our results, we developed a theory of plasmon-assisted energy transfer between SNPs. For NP complexes with relatively large dimensions ($L_{complex} > R_{NP}$), we derived convenient analytical formulas based on the dipole approximation. In the general case, the energy transfer rates are expressed through the response function $F(\omega)$ and should be calculated numerically. We performed our calculations for both Ag- and Au-based nano-complexes. Our theory can be applied to a variety of semiconductor and metal nanocrystal structures.

**Acknowledgements**

This work was supported by NSF and BNNT Initiative at Ohio University.

## Appendix 1

We now apply the equation of motion of the density matrix to the exciton dynamics of SNP2:

$$\frac{\partial \hat{\rho}_{ij}}{\partial t} = \frac{i}{\hbar} \langle i|[\hat{\rho}, \hat{H}]|j \rangle - \sum_{nm} \Gamma_{ij,nm} \hat{\rho}_{nm}, \qquad (A1)$$

where $\hat{\rho}$ and $\hat{H}$ are the density matrix and the Hamiltonian, respectively; the matrix $\Gamma_{ij,nm}$ describes incoherent energy relaxation within the NPs and the index $i$ represents all states of SNP2 ($|exc_2\rangle$). The Hamiltonian is $\hat{H} = \hat{H}_0 + e\Phi_{tot}(r) \cdot e^{-i\omega_1 t} + e\Phi_{tot}^*(r) \cdot e^{i\omega_1 t}$. The current induced in the SNP2 is given by: $\vec{j}_{exc}(t) = \frac{\hbar e}{2im_0} \sum_{m,n} \rho_{mn} \left( \Psi_{exc,m}^* \vec{\nabla} \Psi_{exc,n} - \Psi_{exc,n} \vec{\nabla} \Psi_{exc,m}^* \right)$, where the functions $\Psi_{exc,m}^*$ include the wave functions for the states $|exc_2\rangle$ and for the vacuum state $|0_2\rangle$. In the linear regime, the most important matrix element is given by[22]

$$\rho_{exc_2,0} = ed_{exc} e^{-i\omega_1 t} \int_{SNP2} d^3r \cdot \Psi_{exc_2}^2(r) \frac{\left( \vec{E}_{tot,\beta}(r) \cdot \vec{e}_{exc_2} \right)}{(E_{exc2} - E_1) - i\Gamma_{exc2}}. \qquad (A2).$$

Then, the dissipation rate can be calculated as

$$\int_{SNP2} d^3r \cdot \frac{\vec{j}_{exc}\vec{E}_{tot}^*}{i\omega_1} = -\sum_{exc_2} e^2 d_{exc}^2 \int_{SNP2} d^3r \cdot \Psi_{exc_2}^2(r) \frac{\left(\vec{E}_{tot,\beta}(r) \cdot \vec{e}_{exc_2}\right)}{(E_{exc_2} - E_1) - i\Gamma_{exc2}}. \quad (A3)$$

To obtain the above equation, we used the equality $\langle exc_2|\hat{\vec{v}}|0\rangle = id_{exc}E_{exc2}$, where $\hat{\vec{v}} = \hat{\vec{p}}/m_0$ is the electron velocity operator; we also assumed that $\Gamma_{exc_2} \ll E_{exc_2}$ and $|E_{exc_2} - E_1| \ll E_1$. In addition we took the exciton wavefunctions in the form $|exc_2\rangle = \Psi_{exc_2}(r_e)\Psi_{exc_2}(r_h)u_{exc_2}(\mathbf{r}_h,\mathbf{r}_e)$, where the Bloch function $u_\alpha(\mathbf{r}_h,\mathbf{r}_e)$ has an interband dipole moment parallel to $\vec{e}_{exc_2}$.

## Appendix 2

We now modify slightly eq. 16. This equation describes ET transfer process $|\psi_1\rangle = |1\beta\rangle \to |\chi\rangle$, where $|\chi\rangle$ denotes all states in the system excluding the excitons $|1\beta\rangle$. This approach assumes weak, incoherent coupling of the excitons $|1\beta\rangle$ with the other states. The Hamiltonian of the system reads $\hat{H} = \sum_n \hat{H}_n^0 + \hat{U}_{Coul}$, where $\hat{H}_n^0$ are the Hamiltonians of single NPs and $\hat{U}_{Coul}$ describes Coulomb interaction between the NPs. The off-diagonal components of density matrix can be obtained from the equation

of motion of the density matrix: $\langle 1\beta|\hat{\rho}|\chi\rangle = -\langle 1\beta|\hat{U}_{Coul}|\chi\rangle \cdot \langle \chi|\hat{\rho}|\chi\rangle / [E_\chi - E_1 + i\Gamma_1]$,

where $\Gamma_1$ is the off-diagonal broadening of the excitons $|\psi_1\rangle$ in SNP1. The set of matrix elements $\langle 1\beta|\hat{\rho}|\chi\rangle$ describe energy transfer from SNP1 to MNPs and SNP2. The diagonal matrix element $\langle 1\beta|\hat{\rho}|1\beta\rangle$ can be expressed through $\langle 1\beta|\hat{\rho}|\chi\rangle$. The resultant equation for the diagonal component is:

$$\frac{d\langle 1\beta|\hat{\rho}|1\beta\rangle}{dt} = -(\gamma_{rad,i\alpha} + \gamma^0_{non-rad,i\alpha} + \tilde{\gamma}_{1\beta})\langle 1\beta|\hat{\rho}|1\beta\rangle, \text{ where}$$

$$\tilde{\gamma}_{1\beta} = \tilde{\gamma}_{metal,1\beta} + \tilde{\gamma}_{Forster,1\beta} = \frac{2\pi}{\hbar}\sum_\chi |\langle \chi|\hat{U}_{Coul}|1\beta\rangle|^2 \frac{\Gamma_1}{\pi[(E_1 - E_\chi)^2 + \Gamma_1^2]}. \quad (A4)$$

It is essential to note that this equation is valid for the incoherent regime of ET: $\Gamma_1 \gg \hbar\tilde{\gamma}_{1\beta}$. In other words, the width of exciton states is larger that the parameter $\hbar\tilde{\gamma}_{1\beta}$; this parameter describes the inter-NP interaction. Now we write Eq. A4 in the form of the correlation function (9) with substitution $e^{iE_t t/\hbar} \to e^{iE_t t/\hbar - |t|/\Gamma_1}$. Then, using the inverse Fourier transformation of eq. 9, one can show that

$$\tilde{\gamma}_{1\beta}(E_1) = \int \gamma_{1\beta}(E') \frac{\Gamma_1}{\pi[(E_1 - E')^2 + \Gamma_1^2]} dE'.$$

This equation introduces the broadening due to the donor SNP1 and leads to eq. 17.

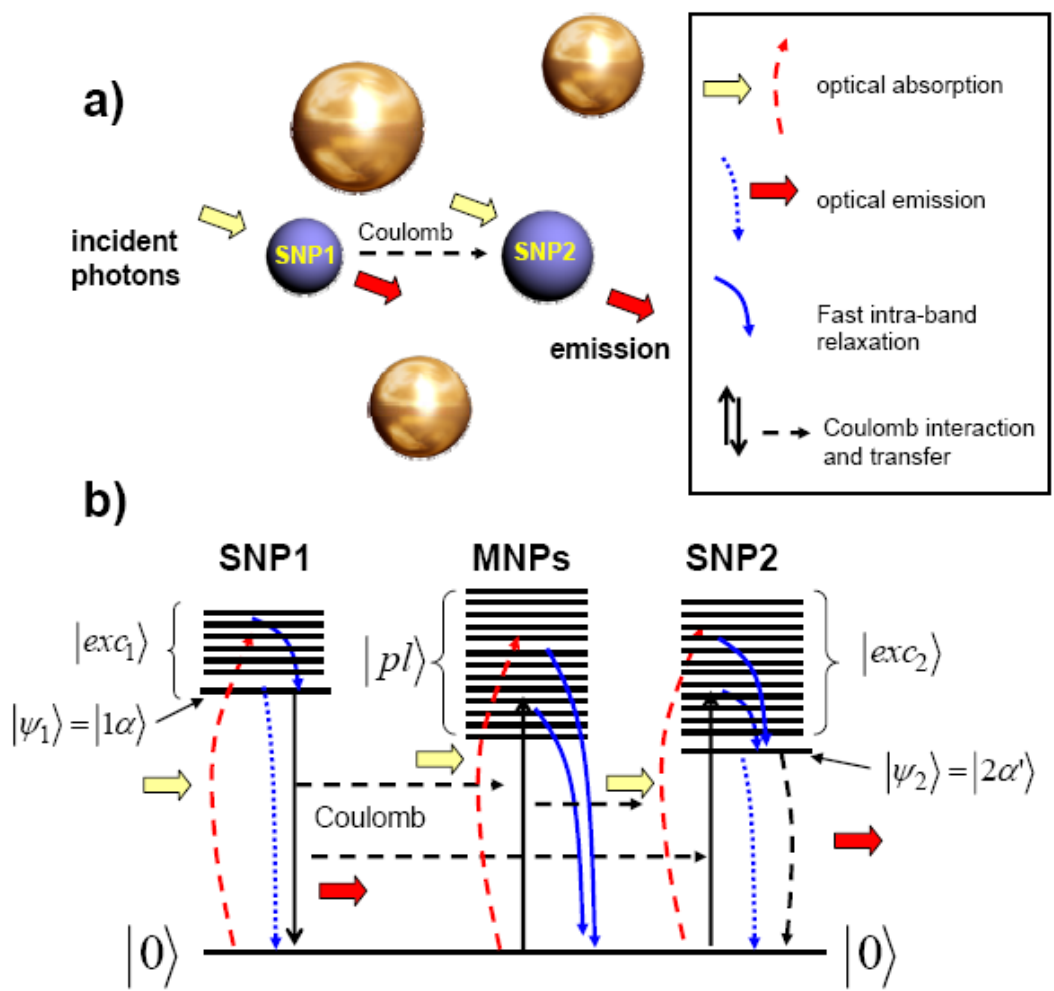

Fig. 1. a) Schematics of the system. b) Energy diagram of FT process and other related processes. The processes are specified in the right upper corner.

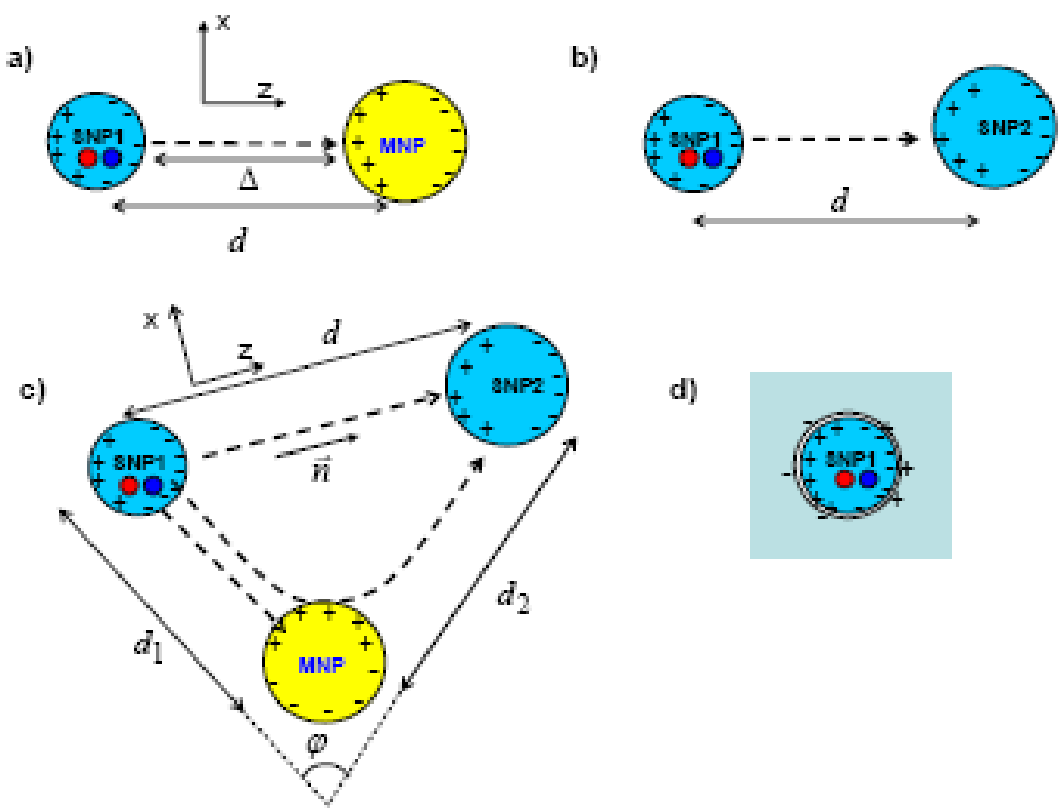

Fig. 2. Schematics of NP complexes. Figures a) , b) and c) depict the processes of energy transfer between NPs. Figure d) illustrates the role of surface changes.

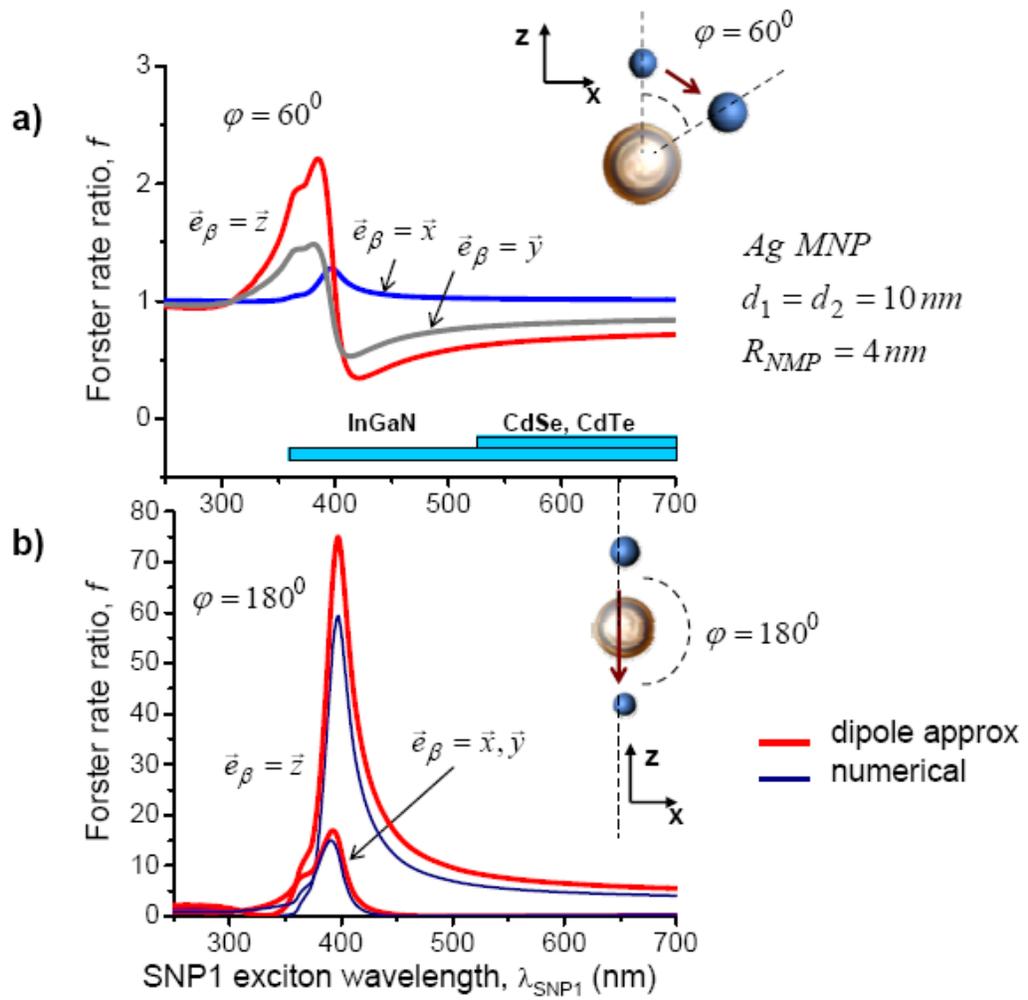

Fig. 3. a) Calculated FT ratios $f_\beta$ for the geometry with $\varphi = 60^0$. b) The same for the geometry $\varphi = 180^0$; red and dark blue curves correspond to the dipole approximation and numerical results. Insets show NP complexes to scale.

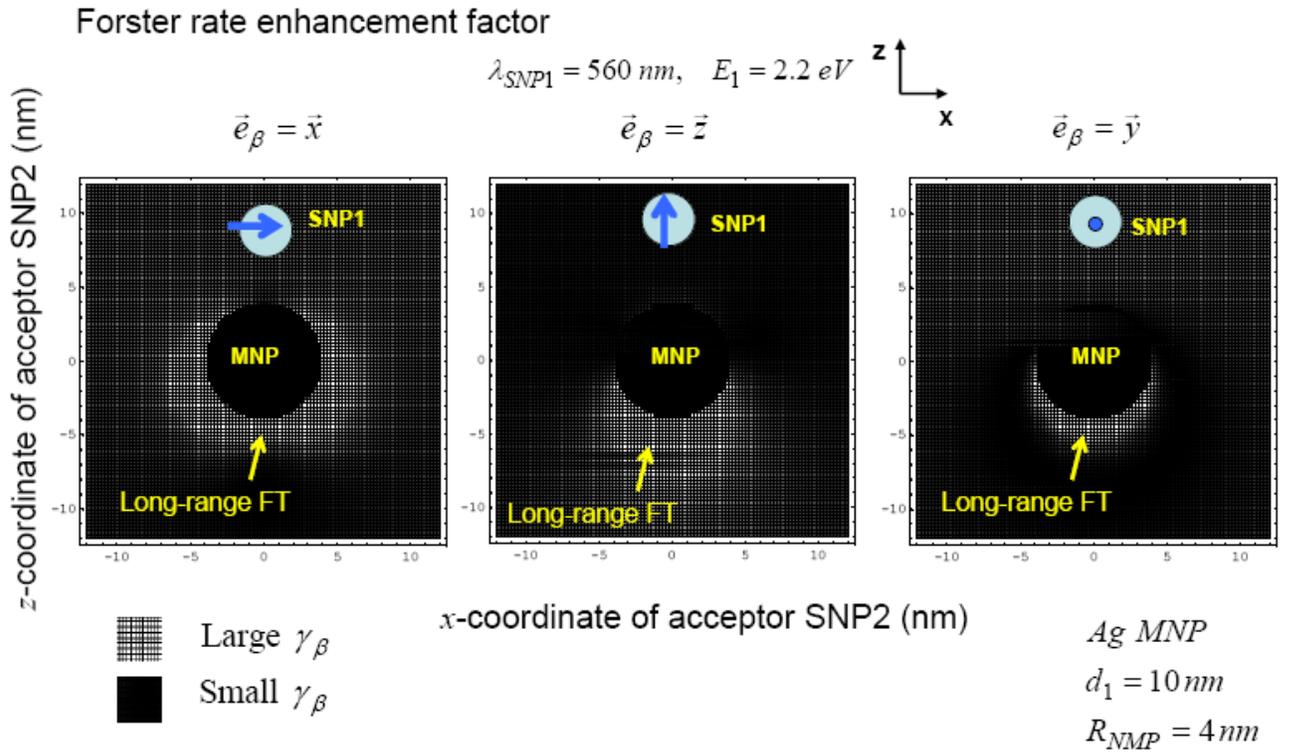

Fig. 4. Grey scale plot of the FT rate ratios $f_\beta$; the position of SNP2, $\vec{r}_{SNP2}$, is a variable. An exciton of SNP1 is transferred to SNP2. Three maps are related to three polarizations of exciton in SNP1; $\lambda_{SNP1} = 560\,nm$.

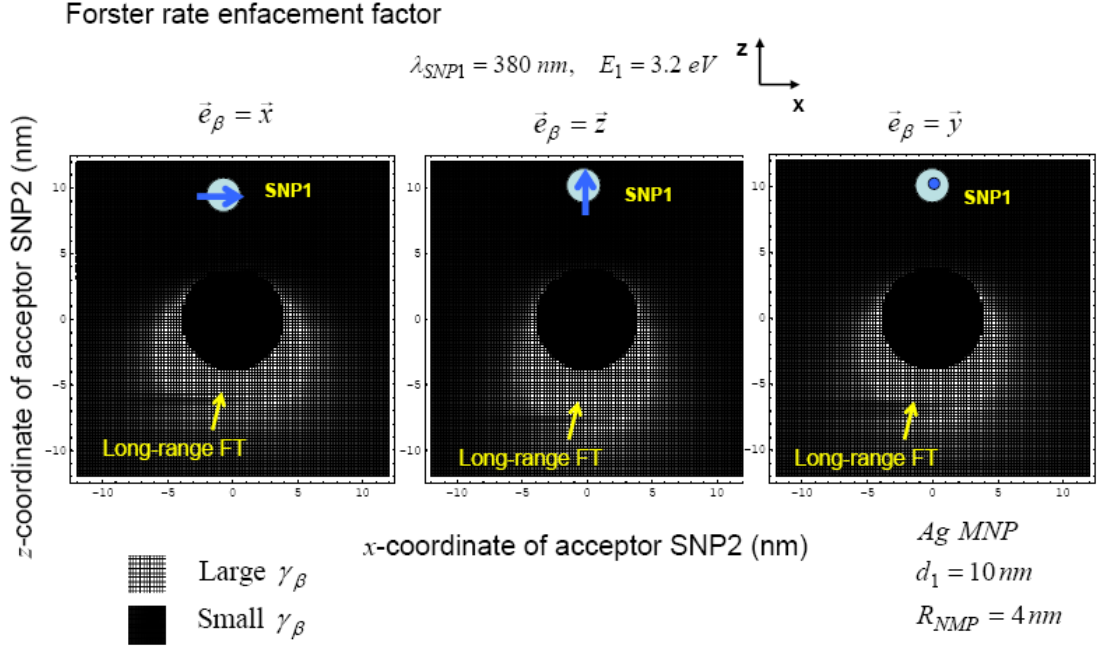

Fig. 5. Grey scale plot of the FT rate ratios $f_\beta$; the position of SNP2, $\vec{r}_{SNP2}$, is a variable. An exciton of SNP1 is transferred to SNP2. Three maps are related to three polarizations of exciton in SNP1; $\lambda_{SNP1} = 380\,nm$.

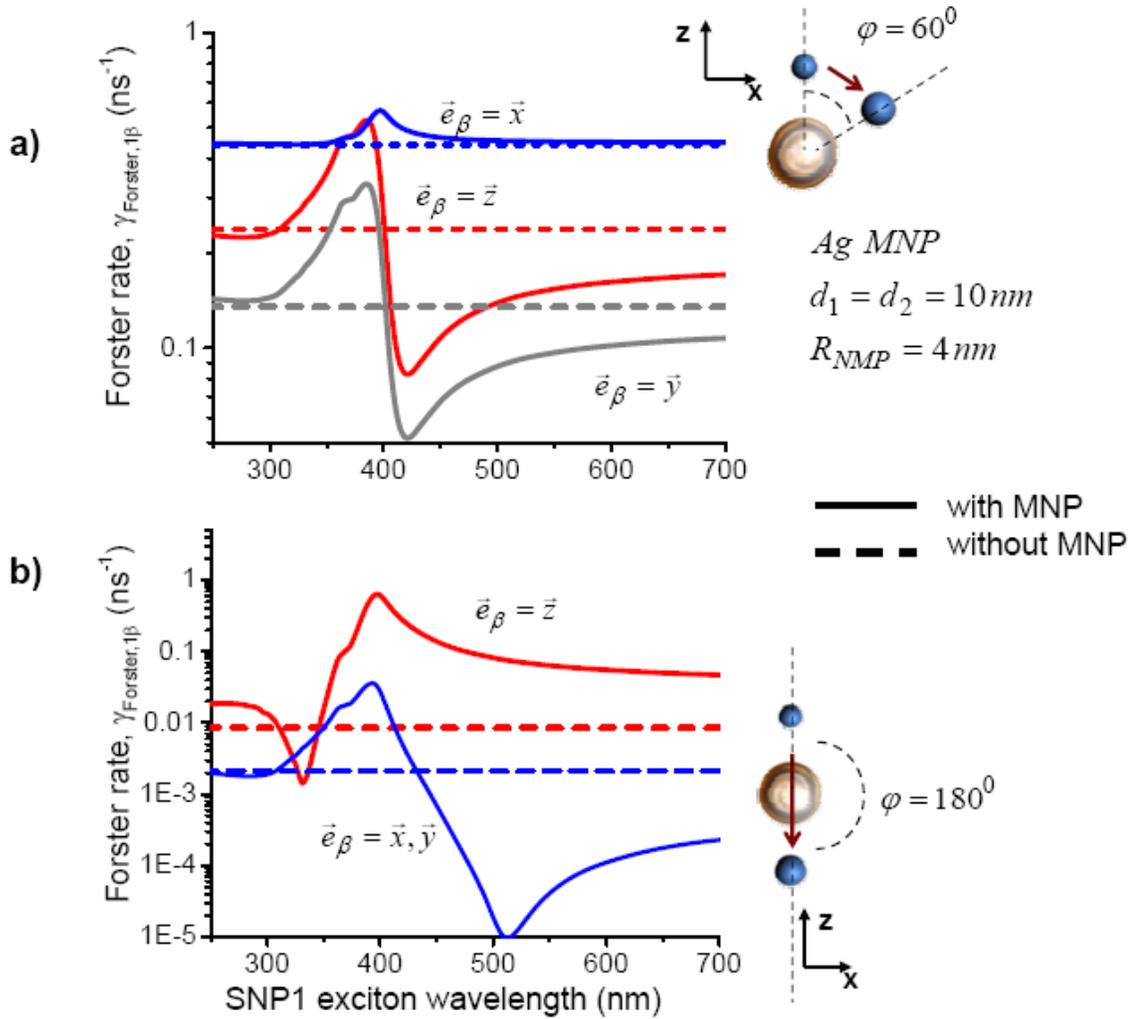

Fig. 6. a) Calculated FT rates $\gamma_{Forster,\beta}$ for two geometries with $\varphi = 60^0$ and $\varphi = 180^0$. Dashed curves are the FT rate in the absence of MNP, whereas solid lines represent the plasmon-assisted FT processes. Insets show NP complexes to scale.

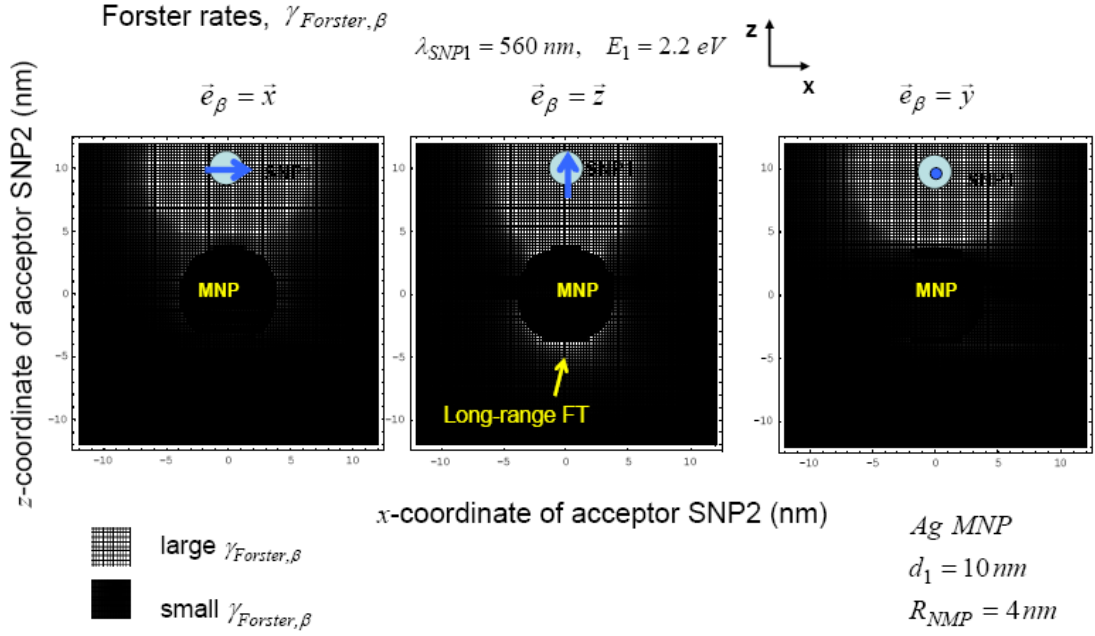

Fig. 7. Grey scale plot of the FT rate $\gamma_{Forster,\beta}$; the position of SNP2, $\vec{r}_{SNP2}$, is a variable. An exciton of SNP1 is transferred to SNP2. Three maps are related to three polarizations of exciton in SNP1; $\lambda_{SNP1} = 560\,nm$.

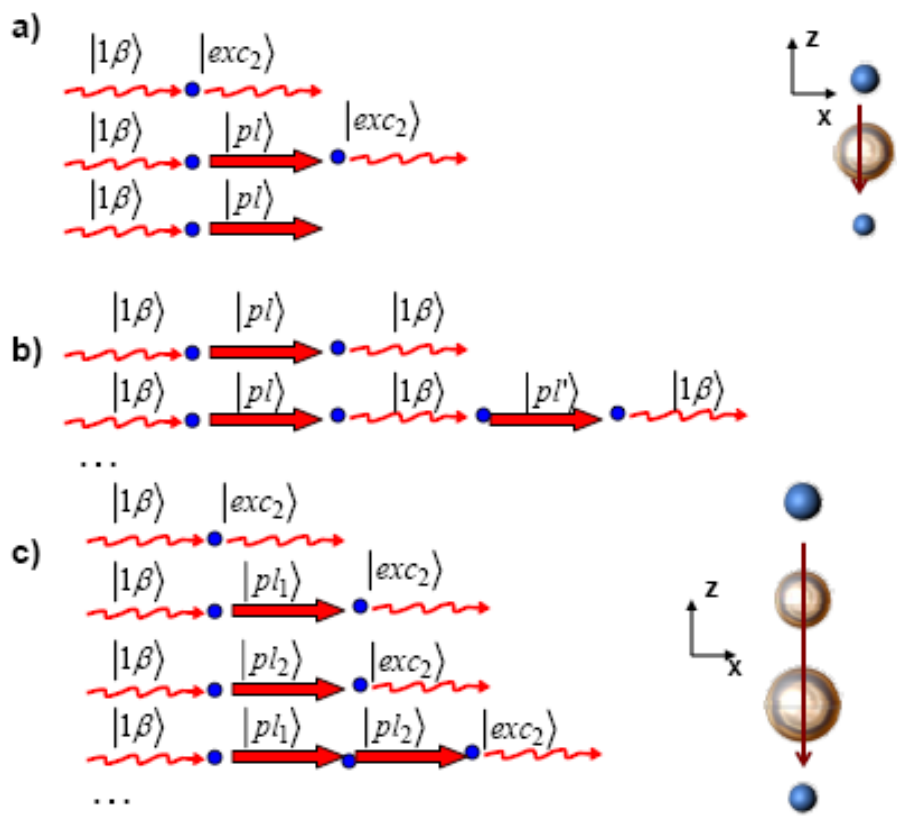

Fig. 8. a) Diagrams responsible for FT, plasmon-assisted FT, and transfer of energy to MNP. b) These diagrams represent the shift of the exciton energy in the presence of plasmons. c) Diagrams for plasmon-assisted FT in the presence of two interacting MNPs.

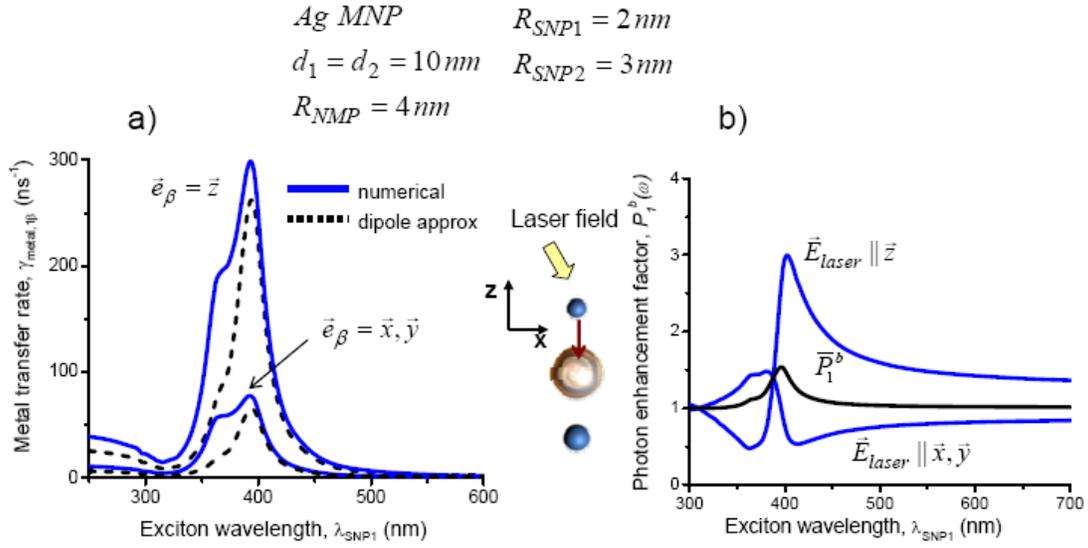

Fig. 9. a) Calculated metal transfer rates $\gamma_{metal,1\beta}$ for the geometries with $\varphi = 180^0$. Dashed curves are obtained within the dipole approximation, whereas solid lines represent numerical results. b) The field enhancement factor for SNP1 calculated numerically.

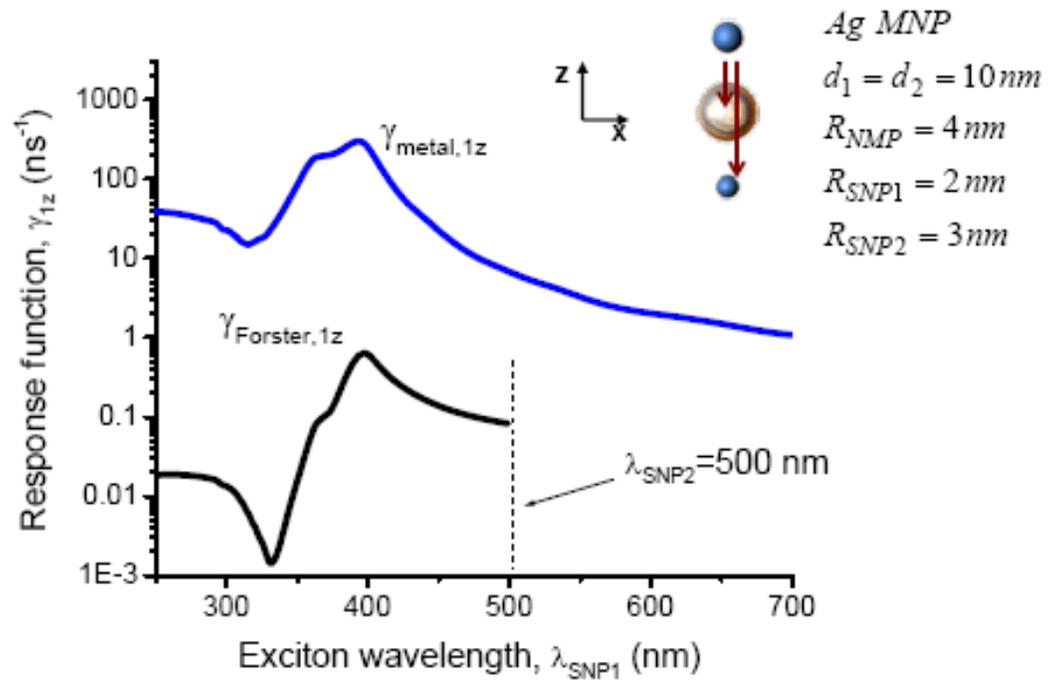

Fig. 10. Calculated components of the response functions, $\gamma_{metal,1z}$ and $\gamma_{Forster,1z}$ for the geometry with $\varphi = 180^0$. The lowest exciton wavelength of SNP2 is assumed to be $\lambda_{SNP2} = 500\,nm$.

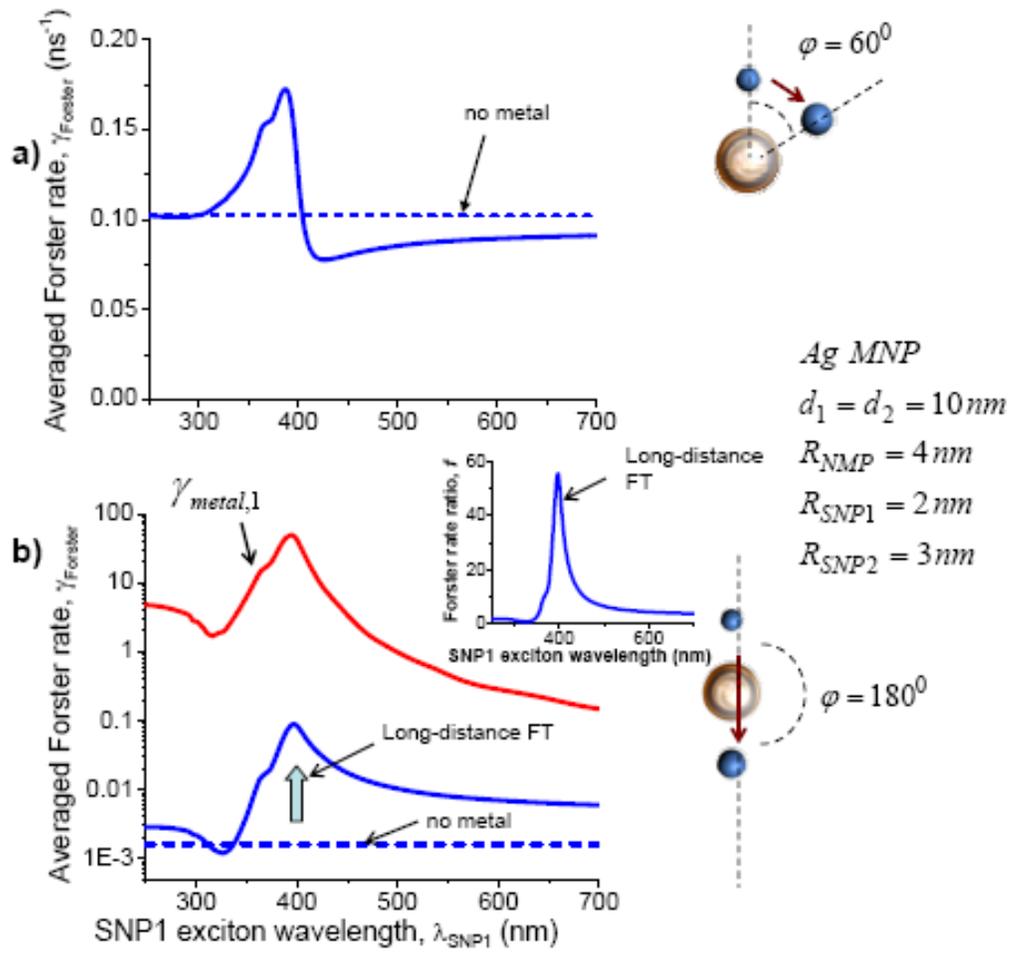

Fig. 11. Calculated averaged FT for two geometries. Inset: FT enhancement factor, $f$, for the second geometry.

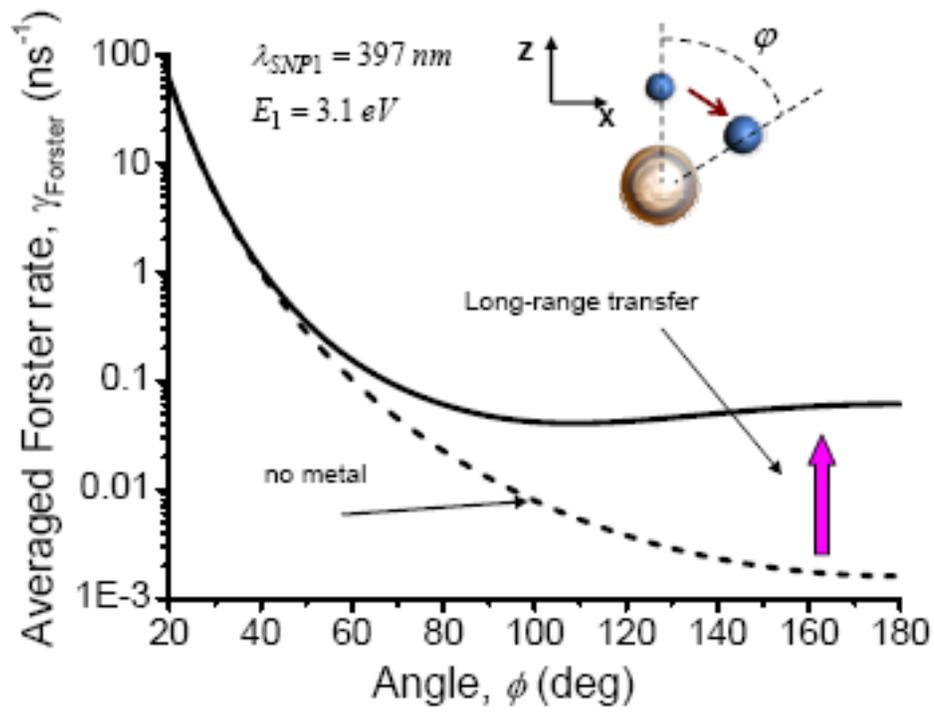

Fig. 12. Calculated averaged FT as a function of the angle $\varphi$. Inset shows the geometry.

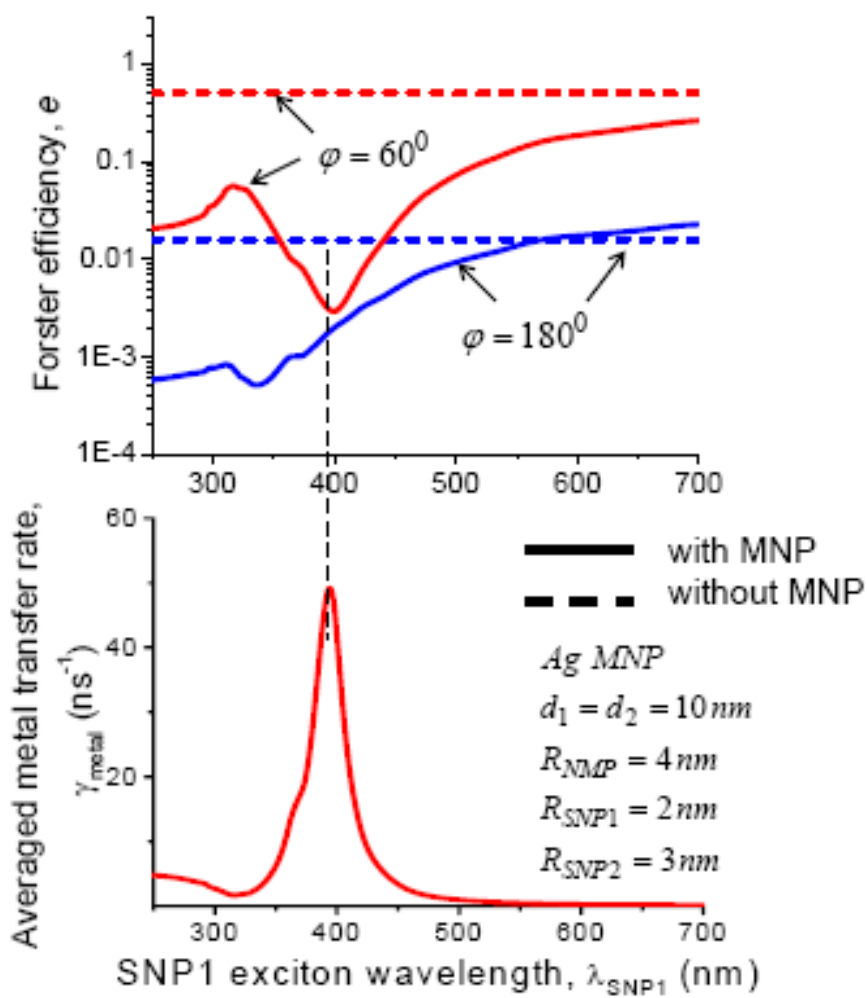

Fig. 13. Calculated efficiency and metal transfer rate as a function of $\lambda_{SNP1}$.

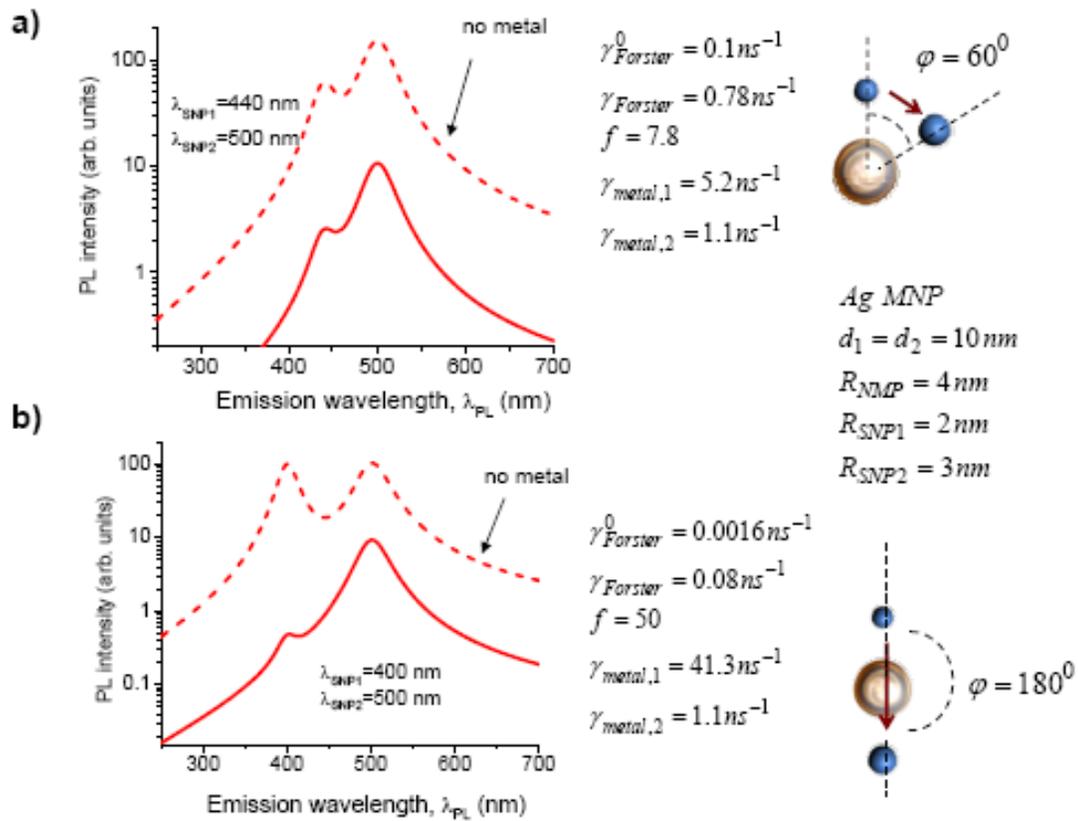

Fig. 14. Calculated PL spectra for two systems. The dashed curves show the spectra in the absence of MNP. Insets: schematics of nano-complexes.

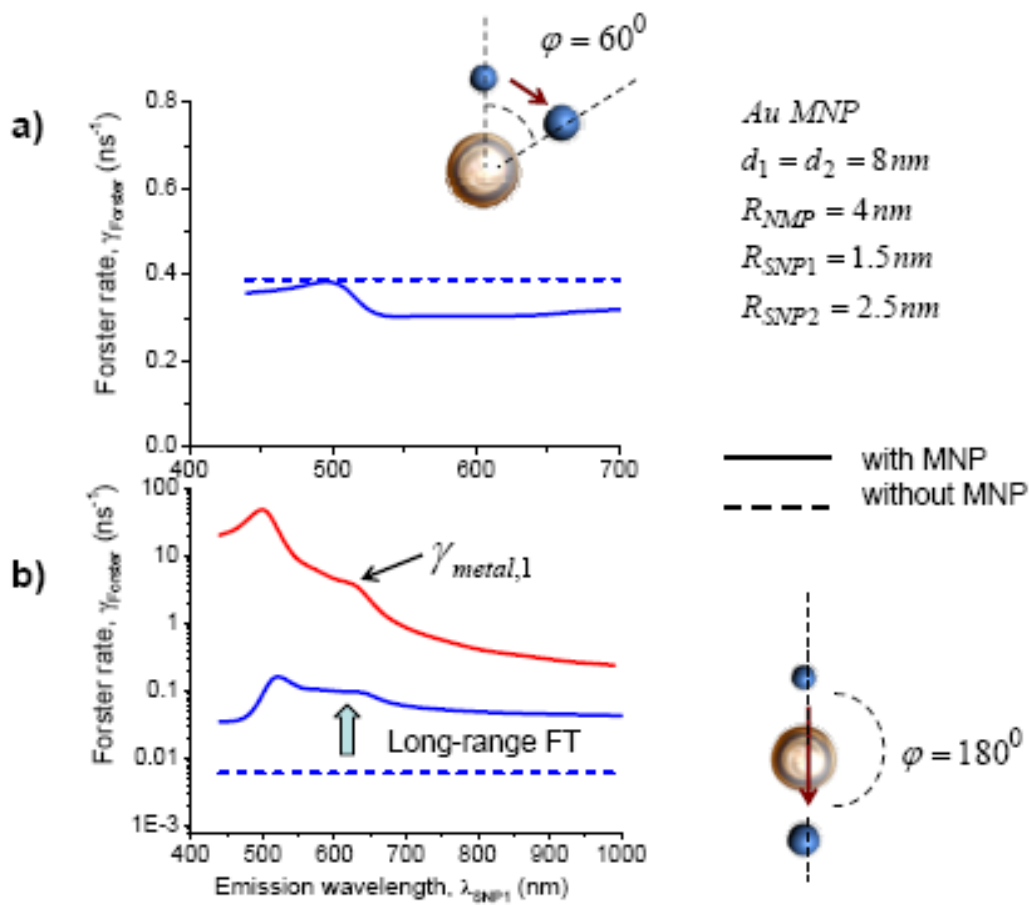

Fig. 15. Calculated averaged FT rates for two complexes incorporating Au-MNP. The panel b) also depicts the metal transfer rate for SNP1. Solid and dashed lines show the results with and without MNP, respectively. Insets: NP complexes.

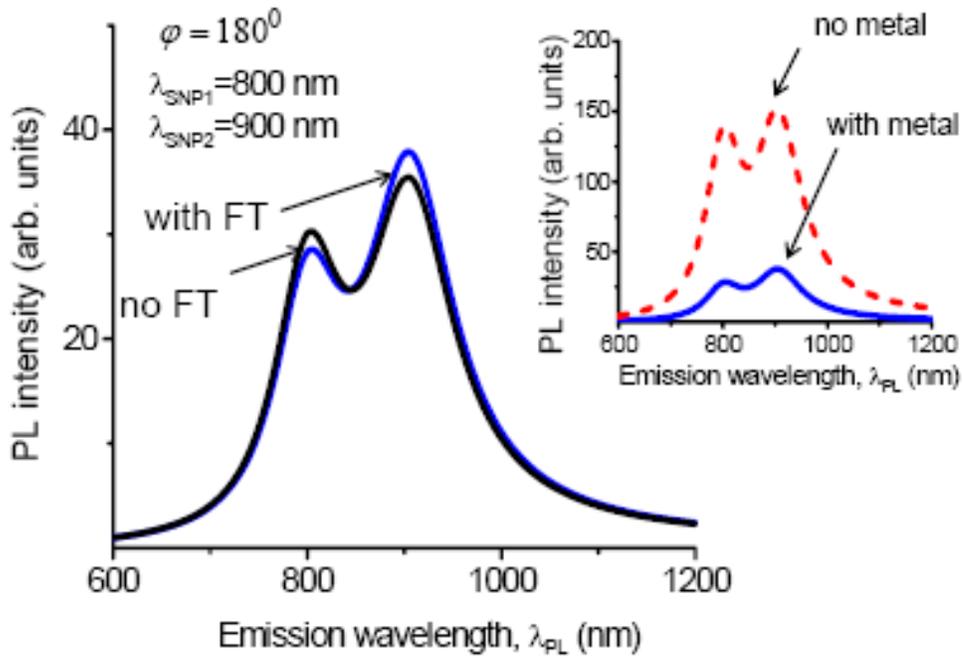

Fig. 16. Calculated PL spectra for a pair of SNPs in the presence of Au-MNP. Two curves show the PL intensity with and without FT. Inset: PL spectrum with and without Au MNP.

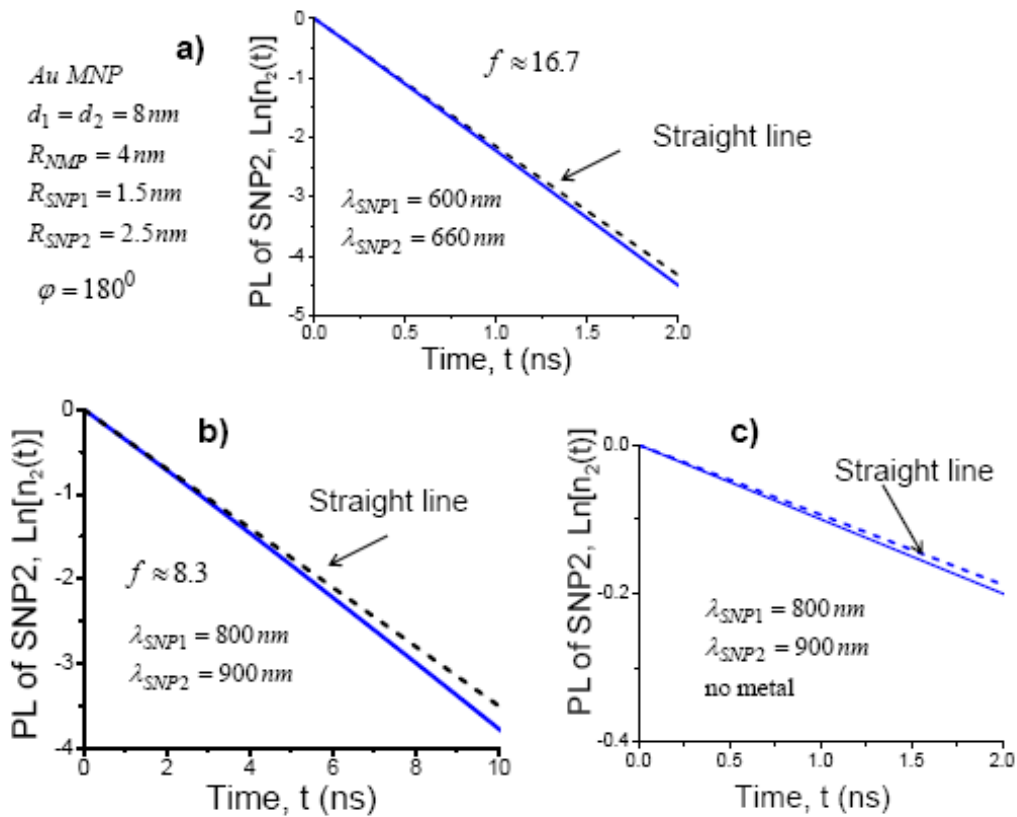

Fig. 17. Temporal dynamics of exciton population of SNP2. Figures a) and b) describe SNP2 exciton dynamics in the presence of Au-NP. Figure c) shows $\ln n_2(t)$ in the absence of MNP. One can see that dynamics in the case c) becomes slower compared to figure b). The straight lines are drawn to show the FT effect. These lines have a slope $d(\ln n_2(t))/dt \,|_{t=0}$.

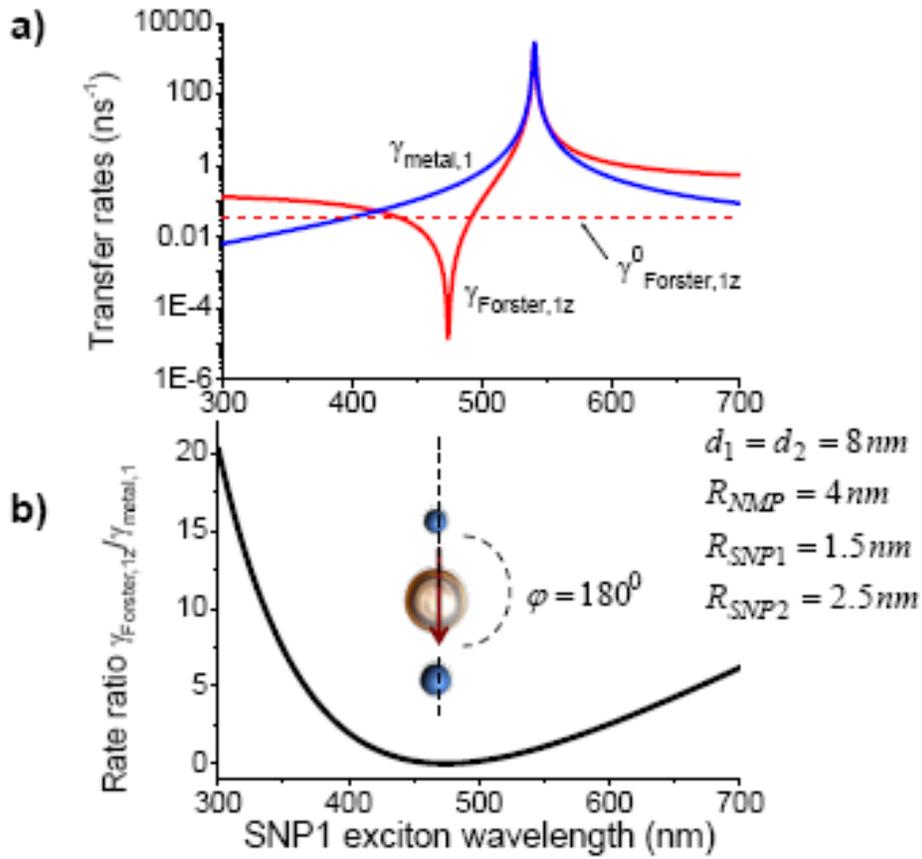

Fig. 18. Calculated FT rate for an idealized model (a). We also show the metal transfer rate (a) and the ratio $\gamma_{Forter,1z}/\gamma_{metal,1}$ (b).